\documentclass[12pt, a4paper]{article}
\pdfoutput=1
\usepackage{graphicx}
\usepackage{amssymb}
\usepackage{amsmath}
\usepackage{bm}
\usepackage[table]{xcolor} %for \rowcolors
\usepackage{cite}
\usepackage{slashed}
\usepackage{subfigure}
\usepackage{epstopdf}            
\usepackage{epsfig}
\usepackage{here}
\usepackage{comment}
\usepackage{booktabs} %for \toprule
\usepackage{colortbl} %for \rowcolor
\usepackage{wrapfig}
\usepackage{ascmac}
\usepackage{fancybox}  
\usepackage{placeins}
\usepackage{url}
\usepackage{bbm}
\renewcommand{\thefootnote}{\fnsymbol{footnote}}
\def\beq#1\eeq{\begin{align}#1\end{align}}
\newcommand{\ov}{\overline}

\newcommand{\eg}{{e.g.}}

\RequirePackage{xspace}
\def\Bbar    {\kern 0.18em\overline{\kern -0.18em B}{}\xspace}

\usepackage{threeparttable}
\usepackage{caption}
\definecolor{BlueViolet}{rgb}{0.2, 0.00, 0.7}
\definecolor{Blue}{rgb}{0.15, 0.00, 0.9}
\definecolor{lightblue}{rgb}{0.15, 0.35, 0.95}
\definecolor{kitgreen}{rgb}{0, 
0.58823 %150/255
, 0.50980 %130/255
}
\usepackage[%dvipdfmx,
colorlinks=true, linkcolor=lightblue,citecolor=lightblue,urlcolor=kitgreen]{hyperref} 

\newcommand{\ttp}{{
    \includegraphics[height=0.8em]{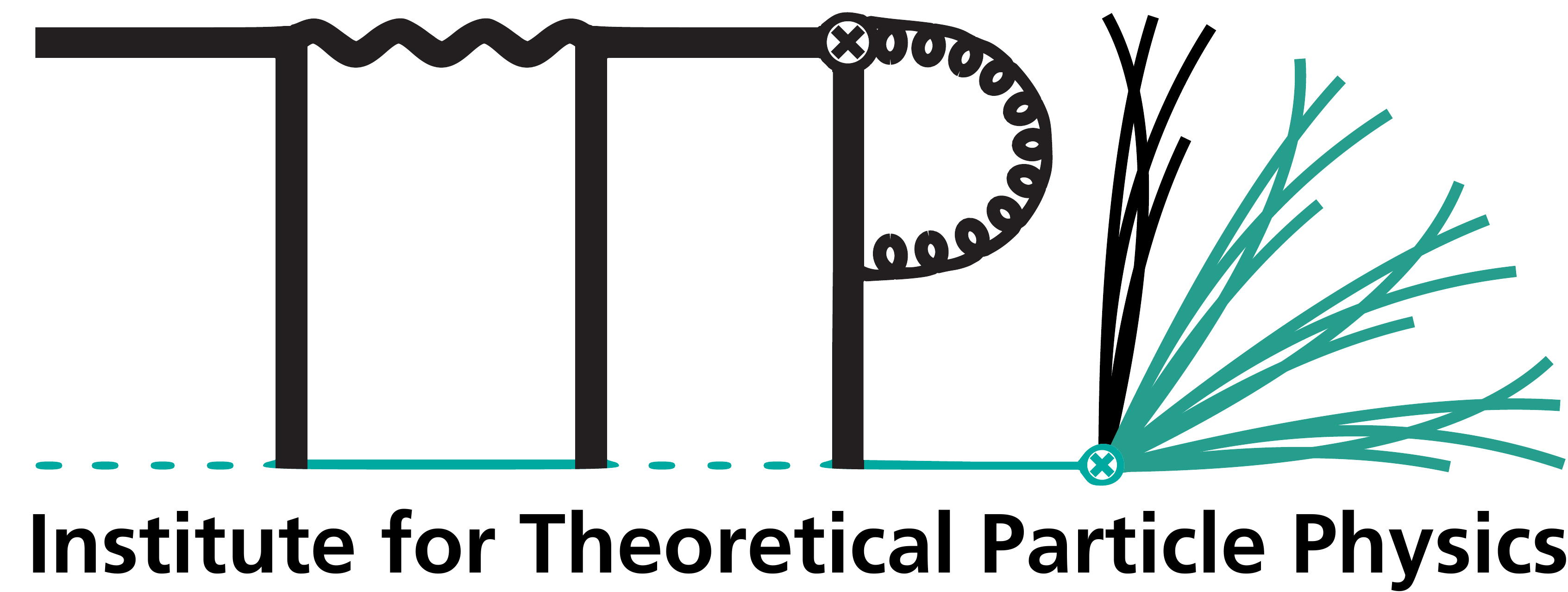}
}}
\newcommand{\iap}{{
    \includegraphics[height=0.8em]{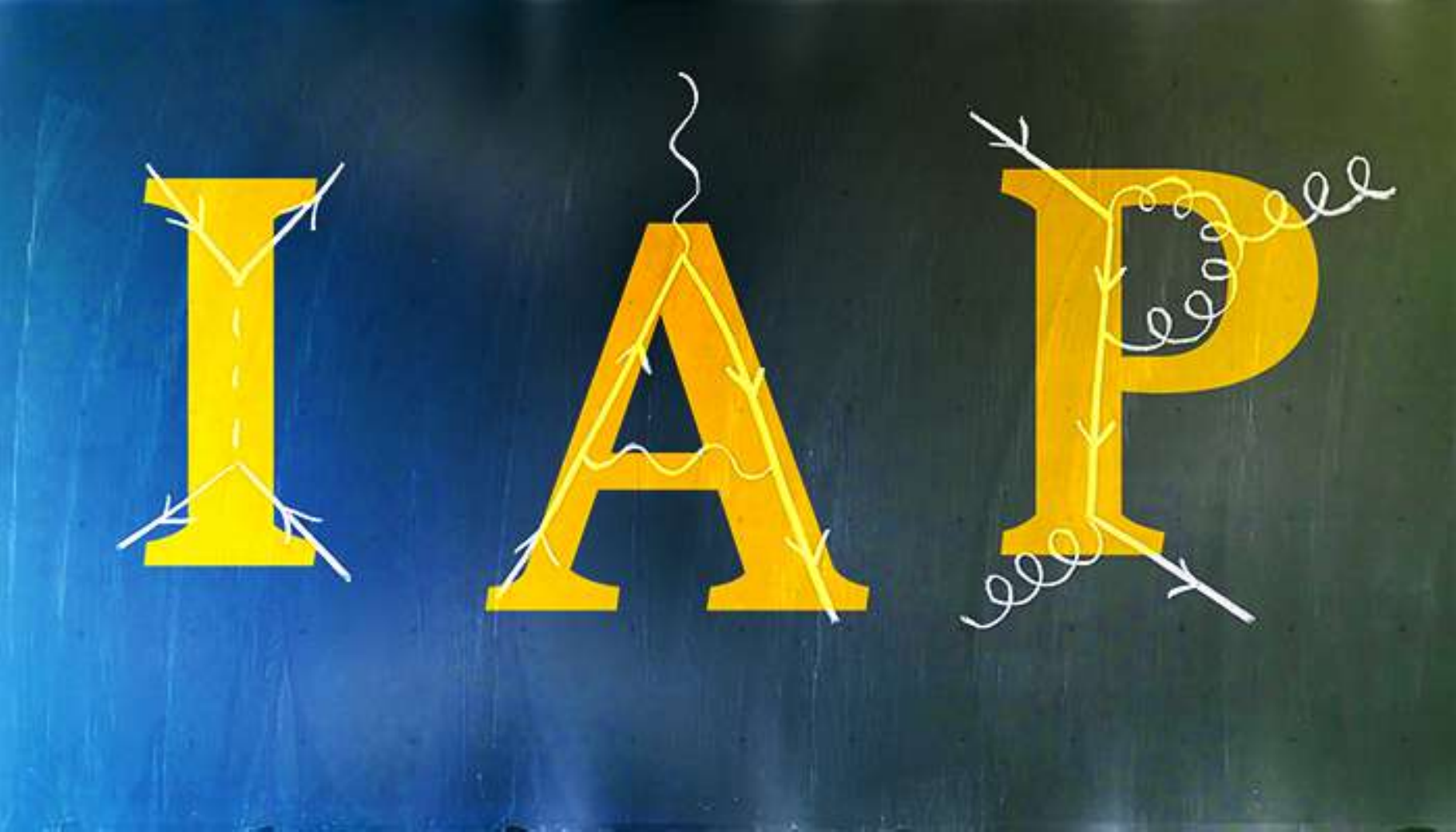}
}}
\newcommand{\nagoya}{{
    \includegraphics[height=0.8em]{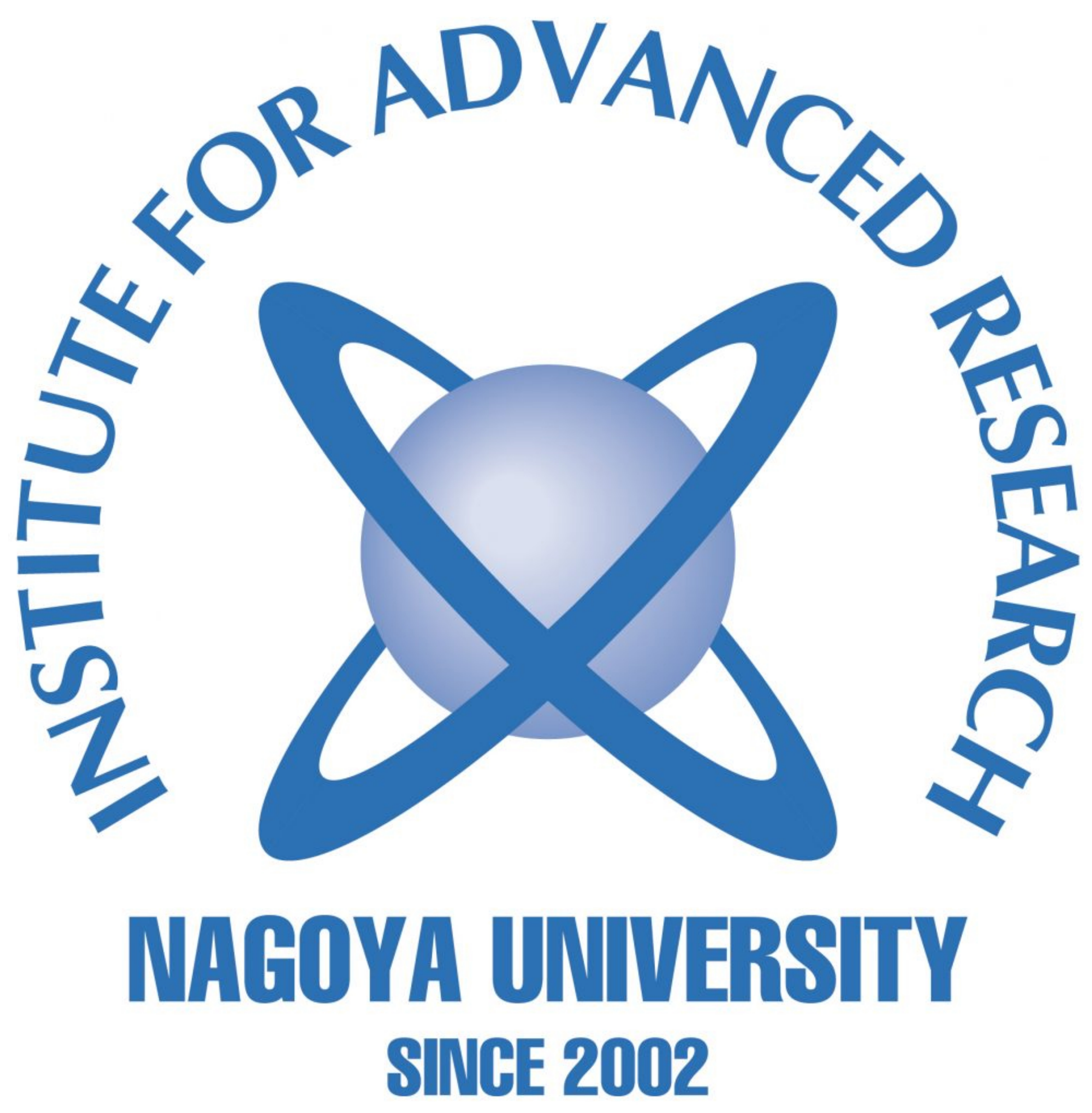}
}}
\newcommand{\kmi}{{
    \includegraphics[height=0.8em]{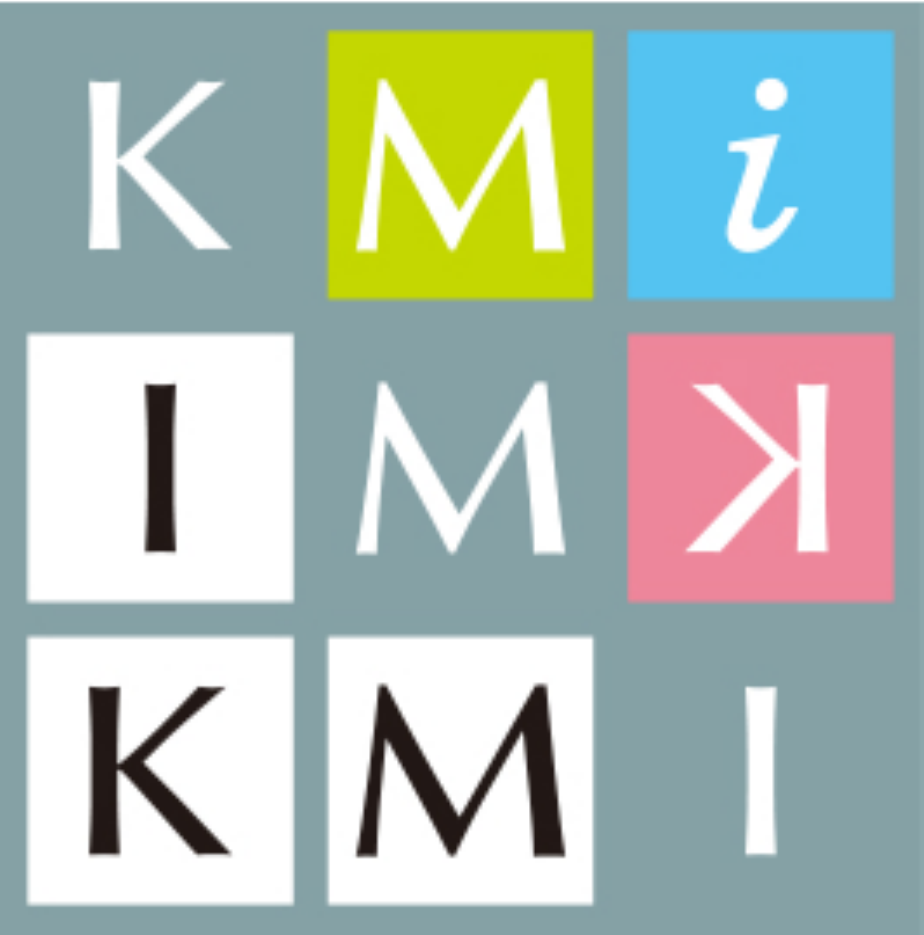}
}}
\newcommand{\kek}{{
    \includegraphics[height=0.8em]{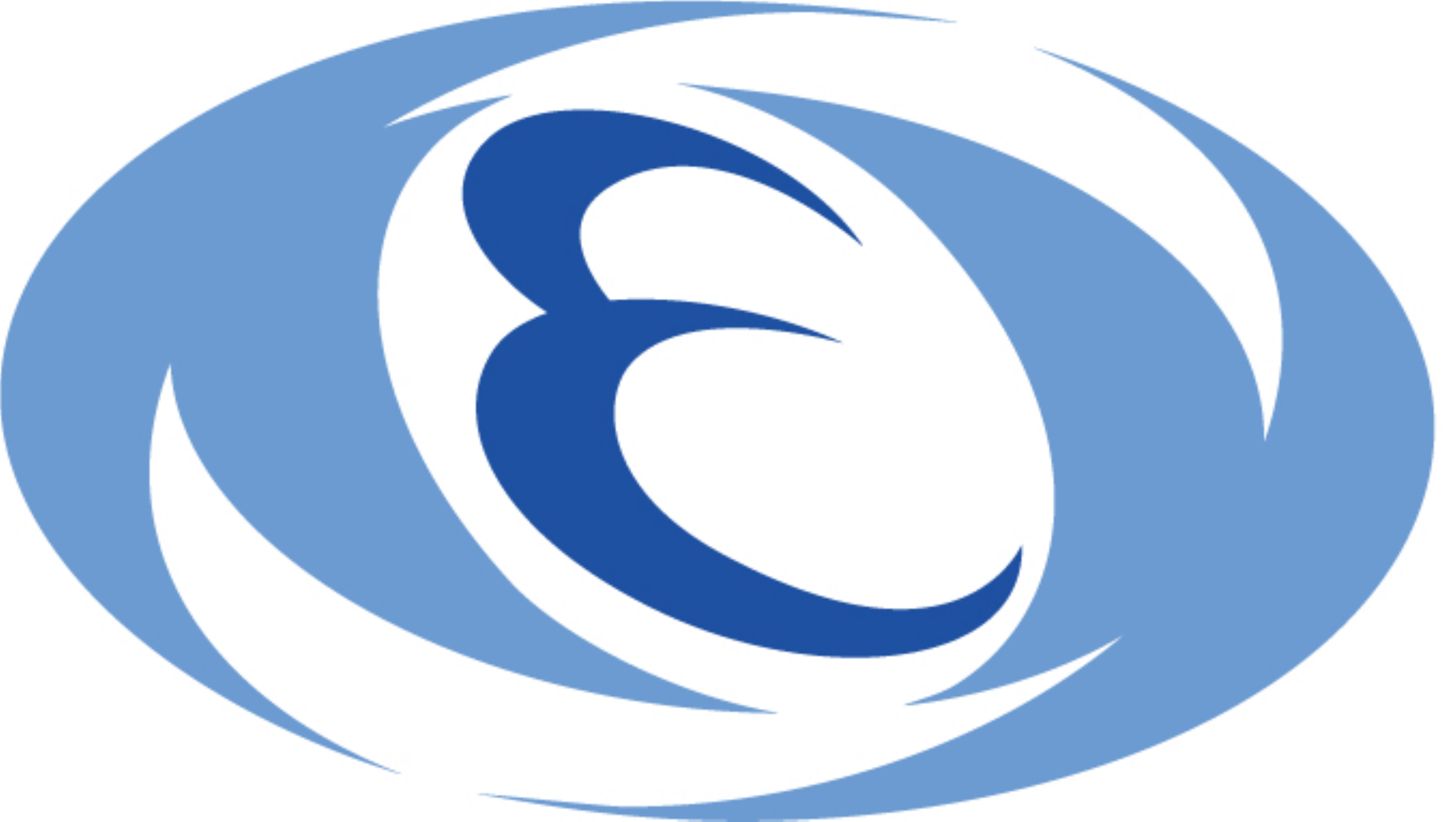}
}}
\newcommand{\itp}{{
    \includegraphics[height=0.8em]{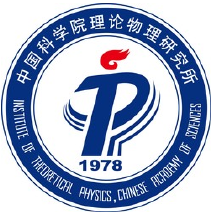}
}}
\newcommand{\kindai}{{
    \includegraphics[height=0.8em]{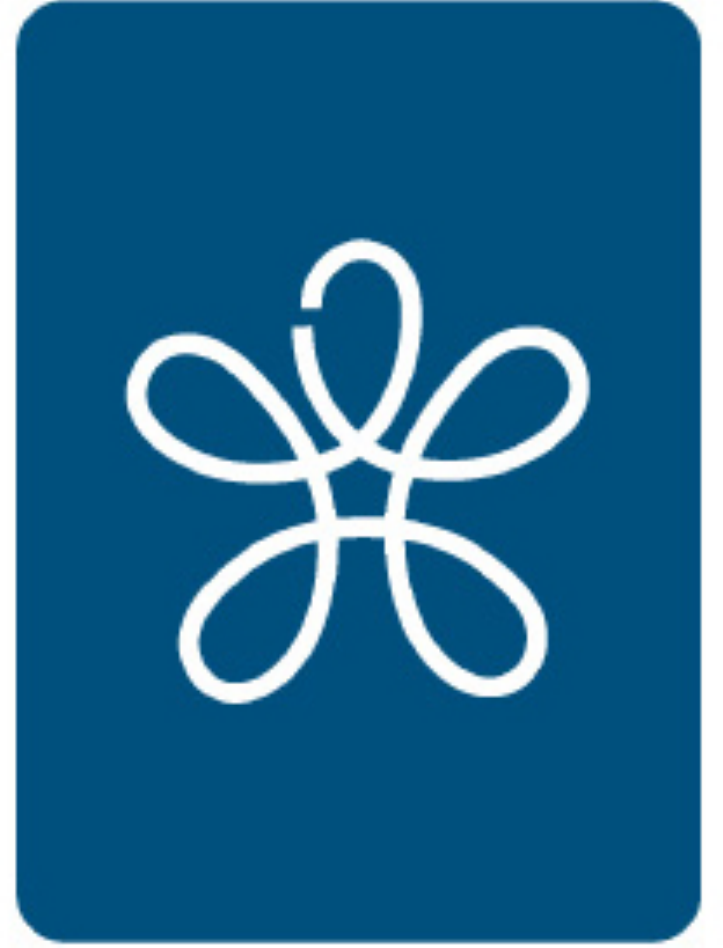}
}}

\begin{document}
\sloppy %https://tex.stackexchange.com/questions/9107/how-can-i-make-my-text-never-go-over-the-right-margin-by-always-hyphenating-or-b

\begin{titlepage}

\begin{center}

\hfill{
P3H--22--102, TTP22--061, KEK--TH--2468}

\vskip .3in

% bold applies to math too
\makeatletter\g@addto@macro\bfseries{\boldmath}\makeatother

{\Large{\bf Chasing the two-Higgs doublet model\\[0.45em] 
in the di-Higgs production}}

\vskip .3in

{\large 
Syuhei Iguro,$^{\ttp,\iap}$
}
{\large 
Teppei Kitahara,$^{\nagoya,\kmi,\kek,\itp}$
} 
{\large
Yuji Omura,$^{\kindai}$\\[0.2em]
and}
{\large
Hantian Zhang$^{\ttp}$}
\vskip .15in

{\small
\begin{tabbing}
$\ttp$ \= {\it 
Institute for Theoretical Particle Physics (TTP), Karlsruhe Institute of Technology (KIT),}\\[0.3em]
\> {\it  Engesserstra{\ss}e 7, 76131 Karlsruhe, Germany}
\\[0.3em]
$~~\iap$ \> {\it Institute for Astroparticle Physics (IAP),
KIT, 
Hermann-von-Helmholtz-Platz 1,}\\[0.3em]
\> {\it 76344 Eggenstein-Leopoldshafen, Germany}
\\[0.3em]
$~~~~\nagoya$ \> {\it 
Institute for Advanced Research, Nagoya University, Nagoya 464--8601, Japan}
\\[0.3em]
$~~~~\kmi$ \> {\it 
Kobayashi-Maskawa Institute for the Origin of Particles and the Universe,} \\[0.3em]
\> {\it  Nagoya University, Nagoya 464--8602, Japan}
\\[0.3em]
$~~\kek$ \> {\it KEK Theory Center, IPNS, KEK, Tsukuba 305--0801, Japan}
\\[0.3em]
$~~~~\itp$ \> {\it CAS Key Laboratory of Theoretical Physics, Institute of Theoretical Physics,} 
\\[0.3em]
\> {\it  Chinese Academy of Sciences, Beijing 100190, China}
\\[0.3em]
$\,~~~~\kindai$ \> {\it Department of Physics, Kindai University, Higashi-Osaka, Osaka 577--8502, Japan}
\\[0.6em]
\> {\it E-mail:} \href{mailto:igurosyuhei@gmail.com}{igurosyuhei@gmail.com},
\href{mailto:teppeik@kmi.nagoya-u.ac.jp}{teppeik@kmi.nagoya-u.ac.jp},\\ ~~~~~~~~~~~~~~~~~
\href{mailto:yomura@phys.kindai.ac.jp}{yomura@phys.kindai.ac.jp},
\href{hantian.zhang@kit.edu}{hantian.zhang@kit.edu}
\end{tabbing}
}

\end{center}

%%%%%%%%%%%%%%%%%%%%%%%%%
\begin{abstract}
\noindent
%%%%%%%%%%%%%%%%%%%%%%%%%
We investigate the di-Higgs production at the Large Hadron Collider in the two-Higgs doublet model (2HDM). 
In particular, we study the production of an extra neutral Higgs boson $\phi$ in association with the Standard Model (SM) Higgs boson $h$ in the Higgs alignment limit.
We analyze two scenarios where the additional Higgs $\phi$ is  CP-even or -odd state with a large top-Yukawa interaction. 
The leading contribution of this production comes from the top-quark loop-induced gluon-fusion channel $gg \to h\phi$.
The measurement of the $h\phi$ production 
can probe the quartic couplings in the Higgs potential 
as well as the top-Yukawa couplings.
Imposing both theoretical constraints (from the perturbative unitarity and the vacuum stability bounds) and experimental bounds (from the SM Higgs and flavor physics measurements) on the 2HDM parameter space,  
we calculate the production cross-section of $gg \to h\phi$.
Furthermore, we scrutinize these processes 
in the parameter spaces where the CMS di-tau and di-photon excesses around 100\,GeV, and/or the muon $g-2$ anomaly can be accommodated.
\end{abstract}
\end{titlepage}

\setcounter{page}{1}
\renewcommand{\thefootnote}{\#\arabic{footnote}}
\setcounter{footnote}{0}

%%%%%%%%%%%%%%%%%%%%%%%%%
% Contents
%%%%%%%%%%%%%%%%%%%%%%%%%
\hrule
\tableofcontents
\vskip .2in
\hrule
\vskip .4in

%%%%%%%%%%%%%%%%%%%%%%%%%

%%%%%%%%%%%%%%%%%%%%%%%%%%%%%%%
\section{Introduction}
\label{sec:Intro}
%%%%%%%%%%%%%%%%%%%%%%%%%%%%%%%
The spontaneous symmetry breaking  of $SU(2)_L\times U(1)_Y$
is caused by a Higgs field in the Standard Model (SM) \cite{Glashow:1961tr,Goldstone:1962es}, where 
the Higgs potential is given by
a negative mass squared term and a quartic coupling, and the Higgs obtains non-vanishing 
vacuum expectation value (VEV).
Through the couplings between the Higgs and the other SM fields, fermions obtain their masses and the $SU(2)_L$ gauge bosons acquire the masses by absorbing the Nambu--Goldstone (NG) bosons \cite{Higgs:1964pj,Englert:1964et,Guralnik:1964eu}.
This picture is very successful in explaining experimental results; however,
the origin of the negative mass squared term is unknown.
Therefore it motivates the further understanding of the Higgs sector. There may be multiple Higgs fields and the scalar potential may be more complicated than the one in the SM.

One promising way to reveal the vacuum structure given by the scalar potential is to test signals involving (extra) scalars in the final state at the Large Hadron Collider (LHC).
The Higgs boson pair production, for instance, gives information about the triple-Higgs coupling~\cite{LHCHiggsCrossSectionWorkingGroup:2016ypw,Cepeda:2019klc}. 
Precise calculations of the dominant contribution $gg\to hh$
within the SM have been performed 
at next-to-leading order (NLO) QCD accuracy \cite{Dawson:1998py,Grigo:2013rya,Borowka:2016ehy,Borowka:2016ypz,Grober:2017uho,Baglio:2018lrj,Davies:2018qvx,Bonciani:2018omm,Xu:2018eos,Davies:2019dfy,Bellafronte:2022jmo}.
Its frontier has been pushing up to  next-to-next-to-next-to-leading order (N$^3$LO)  QCD~\cite{deFlorian:2013jea,deFlorian:2013uza,Grigo:2015dia,Spira:2016zna,Grazzini:2018bsd,Gerlach:2018hen,Banerjee:2018lfq,Chen:2019fhs,Davies:2019xzc,Davies:2021kex,Ajjath:2022kpv} and NLO electroweak (EW)~\cite{Borowka:2018pxx,Davies:2022ram,Muhlleitner:2022ijf} in various approximations.
Currently, it has been still difficult to test the scalar potential parameters directly at the LHC 
due to the small Higgs-pair production cross section compared to the huge QCD background \cite{ATLAS:2008xda,CMS:2008xjf}.
Nevertheless, thanks to the accumulating luminosity at the LHC and improvements in the flavor-tagging algorithm \cite{CMS:2018jrd,ATLAS:2019bwq},
the measurement of the Higgs-pair production has provided profound information on the triple-Higgs coupling
and thus the shape of the Higgs potential
\cite{CMS:2017hea,CMS:2017rpp,ATLAS:2018rnh,ATLAS:2018dpp,ATLAS:2018uni,CMS:2018sxu,ATLAS:2018fpd,ATLAS:2019vwv,CMS:2020tkr,ATLAS:2021ifb,CMS:2022cpr,ATLAS:2022xzm}.
The future prospects are discussed in Refs.~\cite{Cepeda:2019klc,ATLAS:2022qjq}.

Multi-Higgs doublet models often appear as low-energy effective field theories: the supersymmetric standard model \cite{Martin:1997ns,Degrassi:2002fi}, left-right symmetric model \cite{Mohapatra:1974gc}, and so on.
Therefore, it is interesting to 
investigate the
new physics models with extra Higgs doublets.
After the SM Higgs discovery in 2012 \cite{ATLAS:2012yve,CMS:2012qbp}, it has turned out that its
interactions are well consistent with the SM predictions within the current experimental and theoretical uncertainties \cite{ATLAS:2016neq,ATLAS:2022vkf}.
This fact may imply that the additional scalars live in the considerably high energy region which leads to the decoupling feature, or they are hidden from the measurements by the {\em Higgs alignment} feature where the additional Higgs doublets do not mix with the SM-like Higgs doublet.
In this paper, we pursue the latter possibility,
and propose a novel way to probe a relatively light neutral scalar
through the di-Higgs production channel at the LHC.

%%%%%%%%%%%%%%%%%%%
We consider the two-Higgs doublet model (2HDM) as a working example of new physics, and assume the existence of a light additional neutral Higgs boson.
There are many motivations for the light additional scalar.
%%%%%%%%%%%  1stOPT  %%%%%%%%%%%%%%%%%
For example,
it is known that the CP violation of the SM, namely the complex phase of the CKM matrix, is not large enough to generate the observed baryon asymmetry of the Universe (BAU) \cite{Cline:2006ts}.
In addition, the observed 125\,GeV Higgs mass is too heavy for the strong first-order phase transition (SFOPT), which is required  for a viable explanation of the BAU.
For a successful SFOPT, 
the modification of the SM Higgs potential is necessary and hence extensions of the Higgs sector are well motivated \cite{Cline:1995dg,Turok:1990zg,Turok:1991uc,Kanemura:2022ozv}.
It is shown that an additional scalar with $\mathcal{O}(1)$ top-Yukawa coupling can provide the large CP violation and explain the observed BAU \cite{Fuyuto:2017ewj}. 
A joint explanation of the BAU and the radiative neutrino mass with light scalars is also discussed \cite{Ma:2006km,Aoki:2008av,Ahriche:2017iar}.

%%%%%%%%%%%   2HDM g-2 with Barr-Zee diagram  %%%%%%%%%%%%%%
Moreover, a light (pseudo) scalar would explain the discrepancy in the muon anomalous magnetic moment (muon $g-2$) \cite{Aoyama:2020ynm,Muong-2:2006rrc,Muong-2:2001kxu,Muong-2:2021ojo}, whose SM expectation value is based on the
$e^+e^- \to$ hadrons data. Note that
the recent evaluations of a window observable for the hadronic vacuum polarization based on lattice QCD simulation \cite{RBC:2018dos}
are shown in Refs.~\cite{Borsanyi:2020mff,Lehner:2020crt,Wang:2022lkq,Aubin:2022hgm,Ce:2022kxy,Alexandrou:2022amy}, and
a large discrepancy with the $e^+e^-\to \pi^+ \pi^-$ data has been reported in Refs. \cite{Crivellin:2020zul,Keshavarzi:2020bfy,Colangelo:2020lcg}.
In the flavor-conserving 2HDM where only flavor-diagonal Yukawa interactions are introduced,  
an additional light scalar could explain the muon $g-2$ anomaly via the  two-loop Barr-Zee  diagram \cite{Ferreira:2021gke}.
Here, note that a large mass gap between the lightest neutral scalar and the charged scalar is necessary to avoid collider constraints.

%%%%%%%%%%%   low mass resonance %%%%%%%%%%%%%%%%%
In recent years, not significant enough but mild excesses around 95\,GeV have been
reported in $\tau\bar{\tau}$\cite{CMS:2022goy} and $\gamma\gamma$ \cite{CMS:2018cyk} resonance searches  by the CMS collaboration, 
and in $b\bar{b}$ mode by the LEP experiment~\cite{LEPWorkingGroupforHiggsbosonsearches:2003ing}. 
In Ref.~\cite{Iguro:2022dok},
it is shown that 
an additional CP-odd pseudo-scalar $\phi$ with a large $\phi \bar{t}\gamma_5 {t}$ interaction
can still provide a viable solution to the $\tau\bar{\tau}$ excess reported by the CMS, while 
 the CP-even scalar explanation is shown to be difficult.
In the light CP-odd scalar scenario,
the production cross-section of $p p \to t\bar{t}+\tau\bar{\tau}$ 
is suppressed because of the cancellation among diagrams.
We will show that the measurement of the di-Higgs production, $pp \to h \phi$, 
can provide a powerful test for this light CP-odd scenario.

In this paper, motivated by the aforementioned points,
we investigate the impact of $h\phi$ production
at the current and high luminosity (HL) LHC,
where $\phi=H$ and $A$ are CP-even and -odd additional neutral scalars, respectively. 
We show the production cross-sections in a plane of relevant model parameters, and discuss the phenomenological impact 
and relevance to those excesses.
Calculations of various types of Higgs-pair production cross-sections in the relevant beyond SM scenarios are addressed Refs.~\cite{Plehn:1996wb,Djouadi:1999rca,Arhrib:2008pw,Arhrib:2009hc,Dolan:2012ac,Hespel:2014sla,Enberg:2016ygw,Bhattacharya:2019oun,Abouabid:2021yvw,Li:2021cgj,Bahl:2022igd}.
In particular, a comprehensive study
in the gluon-fusion channel 
has been done
based on the benchmark model parameters 
\cite{Abouabid:2021yvw}.

The outline of the paper is given as follows.
In Section~\ref{sec:model},
we introduce the 2HDM and summarize the needed model parameters for our analysis.
In Section~\ref{sec:collider}, we investigate the $h\phi$ production, whose phenomenological impact is given along with relevant constraints from flavor physics and Higgs precision measurements, 
as well as the theoretical constraints from the Higgs potential analysis.
The section~\ref{sec:conclusion} is devoted to the conclusion.

%%%%%%%%%%%%%%%%%%%%%%%%%%%%%%%
\section{Two-Higgs doublet model (2HDM)}
\label{sec:model}
%%%%%%%%%%%%%%%%%%%%%%%%%%%%%%%
We consider the 2HDM where an additional scalar doublet is introduced to the SM.
The general scalar potential of the model is given as
\begin{align}
\begin{aligned}
  V=&M_{11}^2 H_1^\dagger H_1+M_{22}^2 H_2^\dagger H_2-\left(M_{12}^2H_1^\dagger H_2+{\rm h.c.}
  \right)\\
&+\frac{\lambda_1}{2}(H_1^\dagger H_1)^2+\frac{\lambda_2}{2}(H_2^\dagger H_2)^2+\lambda_3(H_1^\dagger H_1)(H_2^\dagger H_2)
+\lambda_4 (H_1^\dagger H_2)(H_2^\dagger H_1) \\
&+
\left\{\frac{\lambda_5}{2}(H_1^\dagger H_2)^2+\left[
\lambda_6 (H_1^\dagger H_1)+\lambda_7 (H_2^\dagger H_2)\right] (H_1^\dagger H_2)+{\rm h.c.}\right\}\,.
   \label{Eq:potential}
   \end{aligned}
\end{align}
Here, we  work in the {\em Higgs basis} where only one doublet takes a VEV \cite{Georgi:1978ri, Donoghue:1978cj, Davidson:2005cw} at the renormalization scale $\mu = m_W$:
\begin{align}
H_1 = \begin{pmatrix} G^+ \\ \frac{1}{\sqrt{2}} (v+h+iG^0) \end{pmatrix} \,,\quad 
H_2 = \begin{pmatrix} H^+ \\ \frac{1}{\sqrt{2}} (H+iA) \end{pmatrix} \,,
 \label{Eq:basis}
\end{align}
where $v\simeq 246\,$GeV and $G$ denotes the NG bosons.\footnote{%
Parameter relations between the 2HDM in the Higgs basis and a general basis (with and without softly-broken $\mathbbm{Z}_2$ symmetry) are summarized in Appendix~\ref{sec:parameters}.
}
The stationary conditions for Eq.~\eqref{Eq:basis} are
\begin{align}
    M_{11}^2 = - \frac{\lambda_1}{2} v^2\,, \quad 
    M_{12}^2 = \frac{\lambda_6}{2} v^2\,.
\end{align}
For simplicity, we further assume the CP-conserving scalar potential,
so that one can define the CP-even and -odd scalar mass eigenstates.
The SM-like Higgs is $h$ and the charged scalar is $H^+$, while 
$H$ and $A$ correspond to additional the CP-even and -odd neutral scalars.

It is noted that $h$-$H$ mixing in their mass basis is suppressed as far as $\lambda_6$ is negligible, and then all $h$ interactions become the same as the SM Higgs boson at the renormalization scale $\mu=m_W$. This condition is known as the Higgs alignment limit  \cite{Gunion:2002zf,Craig:2013hca,Carena:2013ooa,BhupalDev:2014bir}.
In the following analysis, we assume $\lambda_6=0$ and consider  the phenomenology of the scalar bosons in the Higgs alignment limit. In addition, $\lambda_7$ leads to only trilinear couplings of the extra scalars as far as $\lambda_6$ is vanishing, so that
$hH/hA$ production we investigate is independent of the value of $\lambda_7$.

In the Higgs alignment limit ($\lambda_6 \to 0$), scalar masses are related as
%%%%%%%%%%%%%%%%%%%%%%%%%%%%%%%%%%%%%%%
\begin{align}
\begin{aligned}
m_h^2& = \lambda_1 v^2\,,\\
m_{H^\pm}^2 & =M_{22}^2+\frac{\lambda_{3}}{2} v^2\,, \\
m_H^2&= M_{22}^2+\frac{\lambda_{hHH}}{2} v^2\,,\\
m_A^2  &= M_{22}^2+\frac{\lambda_{hAA}}{2} v^2\,,
\end{aligned}
\end{align}
%%%%%%%%%%%%%%%%%%%%%%%%%%%%%%%%%%%%%%%
and the first equation requires $\lambda_1 \simeq 0.26$ to obtain the observed Higgs mass.
For the latter convenience, we also define 
\begin{align}
\label{eq:hHHhAA}
\lambda_{hHH}=\lambda_3+\lambda_4+\lambda_5 \,, \quad
\lambda_{hAA}=\lambda_3+\lambda_4-\lambda_5\,.
\end{align}
Note that the equations above hold regardless of whether the softly-broken $\mathbbm{Z}_2$ symmetry is imposed or not.

The breakdown of the custodial symmetry is stringently constrained by the measurements of the oblique parameters \cite{Peskin:1990zt,Peskin:1991sw,PDG2022}.
The constraints require  $m_H \simeq m_{H^\pm}$ or $m_A \simeq m_{H^\pm}$ for the model
to be consistent with the data \cite{Gerard:2007kn}.
This condition is equivalent to $\lambda_4\simeq - \lambda_5$ or  $\lambda_4\simeq \lambda_5$, respectively, in the Higgs alignment limit.
It can be seen from Eq.~(\ref{eq:hHHhAA}) that the $hHH$ and $hAA$ couplings are controlled by $\lambda_3$, $\lambda_4$ and $\lambda_5$.
The important point here is the sign in front of $\lambda_5$.
Since the $\lambda_5$ term in the potential is proportional to $(H_1^\dagger H_2)^2 + {\rm h.c.}$, there is sign difference between the $hHH$ and $hAA$ couplings.
It is also noted that the $hH^+ H^-$ interaction is controlled only by  $\lambda_3$ in the Higgs alignment limit.

We are interested in the di-Higgs production in the Higgs alignment limit.
We define the Yukawa couplings involving extra scalars as follows \cite{Davidson:2005cw}:
\begin{align}
-\mathcal{L}_{H_2\text{-Yukawa}} =
\ov{Q}^i \left(V_{\rm CKM}^\dag \right)^{ij} \widetilde{H}_2 \rho_{jk}^{u} u_R^k 
+ \ov{Q}^i  H_2 \rho_{ij}^{d} d_R^j  + \text{h.c.}  \,,
\label{eq:Lintgenenral}
\end{align}
with $\widetilde{H}_2 = i \tau_2 H_2^\ast$,
and particularly the $ \rho_{tt}^{u} \equiv \rho_{tt}^{\phi}$ interaction is given as
\begin{align}
-\mathcal{L}_{\rm Yukawa}^t = 
\frac{\rho_{tt}^\phi}{\sqrt{2}}\,\bar{t} H t
- i \frac{\rho_{tt}^\phi}{\sqrt{2}}\,\bar{t} A \gamma_5 t -\left[\rho_{tt}^\phi \ov{t}_R H^+  \left(V_{\rm CKM} d_L\right)_3  + \text{h.c.}\right]
\,.
\label{Eq:Lint}
\end{align}
We work on the basis  where the left-handed quarks are represented as 
$Q^T = \begin{pmatrix} V_{\rm CKM}^\dag u_L \,,  & d_L \end{pmatrix}$ with the mass eigenstates $u_L$ and $ d_L$.
Other Yukawa couplings, for example the tau-Yukawa coupling, will be introduced in the next section. 
The large additional top-Yukawa coupling can be realized in many types of 2HDMs; for example, the type-II 2HDM and a flavor-aligned 2HDM. The decay models of extra scalars depend on the setup. In this paper, we are interested in $hH/hA$ productions where the extra scalars dominantly decay to $\tau \tau$, so that we consider a 2HDM with a large $\rho_{tt}^\phi$ and a sizable $\tau \tau$ coupling, $(\rho_{\tau \tau}^\phi/\sqrt{2})\,\bar{\tau} H \tau + i (\rho_{\tau \tau}^\phi/\sqrt{2})\,\bar{\tau} A \gamma_5 \tau$. The other Yukawa couplings are assumed to be vanishing or relatively small to avoid the constraints from, especially, flavor physics.\footnote{This kind of setup can be realized by the softly-broken $\mathbbm{Z}_2$ symmetric 2HDM, see Ref.~\cite{Iguro:2022dok} and references therein.}

In general $\rho_{tt}^\phi$ can be a complex value,
whereas the collider phenomenology that we are interested in does not change in the presence of the complex phase in the Higgs alignment limit.
In this paper, we take $\rho_{tt}^\phi$ to be real for simplicity.
It is noted that if the additional bottom-Yukawa coupling is sizable, 
the chirality-enhanced amplitude significantly affects $b\to s\gamma$,
and this scenario would be excluded easily.
In the following,
we label the additional lighter  and heavier neutral scalars as $\phi_l$ and $\phi_h$, respectively.

%%%%%%%%%%%%%%%%%%%%%%%%%%%%%%%%%%%%%%%%%%%%%%%%%
\section{Phenomenology}
\label{sec:collider}
%%%%%%%%%%%%%%%%%%%%%%%%%%%%%%%%%%%%%%%%%%%%%%%%% 
In this section, we consider the $h\phi_l$ production at the LHC.
We evaluate the production cross section in Sec.~\ref{sec:hphi_production} and discuss the phenomenological constraint on the relevant top-Yukawa coupling in Sec.~\ref{sec:const_ptt}.
The phenomenological impacts on the excesses are investigated in Sec.~\ref{sec:impact}.

%%%%%%%%%%%%%%%%%%%%%%%%%%%%%%%%%%%%%%%%%%%%%%%%%
\subsection{\texorpdfstring{$h \phi_l$}{h phi l} production}
\label{sec:hphi_production}
%%%%%%%%%%%%%%%%%%%%%%%%%%%%%%%%%%%%%%%%%%%%%%%%% 
In the SM,
it is known that there is a partial cancellation among the diagrams in the Higgs pair production in the gluon-fusion processes  \cite{LHCHiggsCrossSectionWorkingGroup:2016ypw,Cepeda:2019klc}.
At the leading order, there are two types of diagrams:
one is the top box diagram and the other is the top triangle diagram with the triple Higgs coupling.
Therefore,
the modified triple Higgs coupling can be probed in this channel, and this 
process has eagerly been studied to test the structure of the Higgs potential.
%%%%%%%%%%%%%%%%%%%%%%%%%%%%%%%%%%%%%%%%%%%%%%%%%
\begin{figure}[t]
\begin{center}
\includegraphics[scale=0.45]{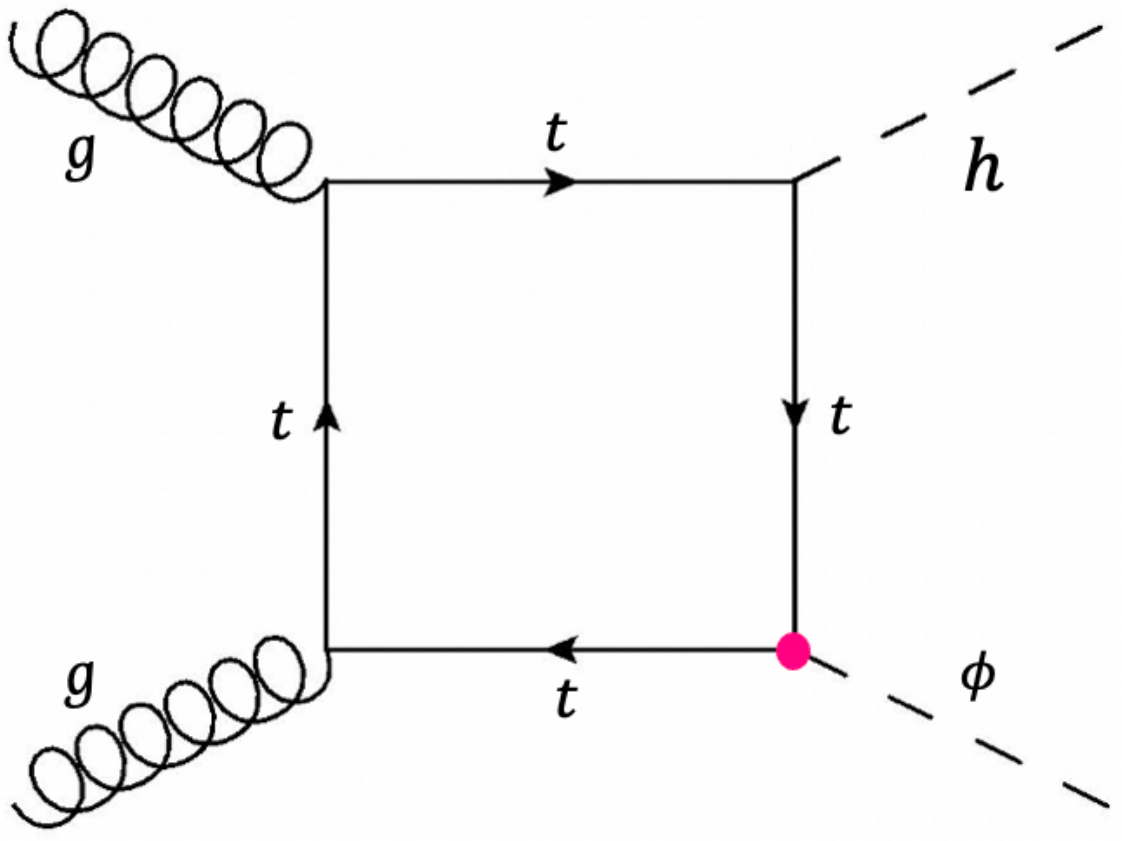} \qquad \qquad 
\includegraphics[scale=0.45]{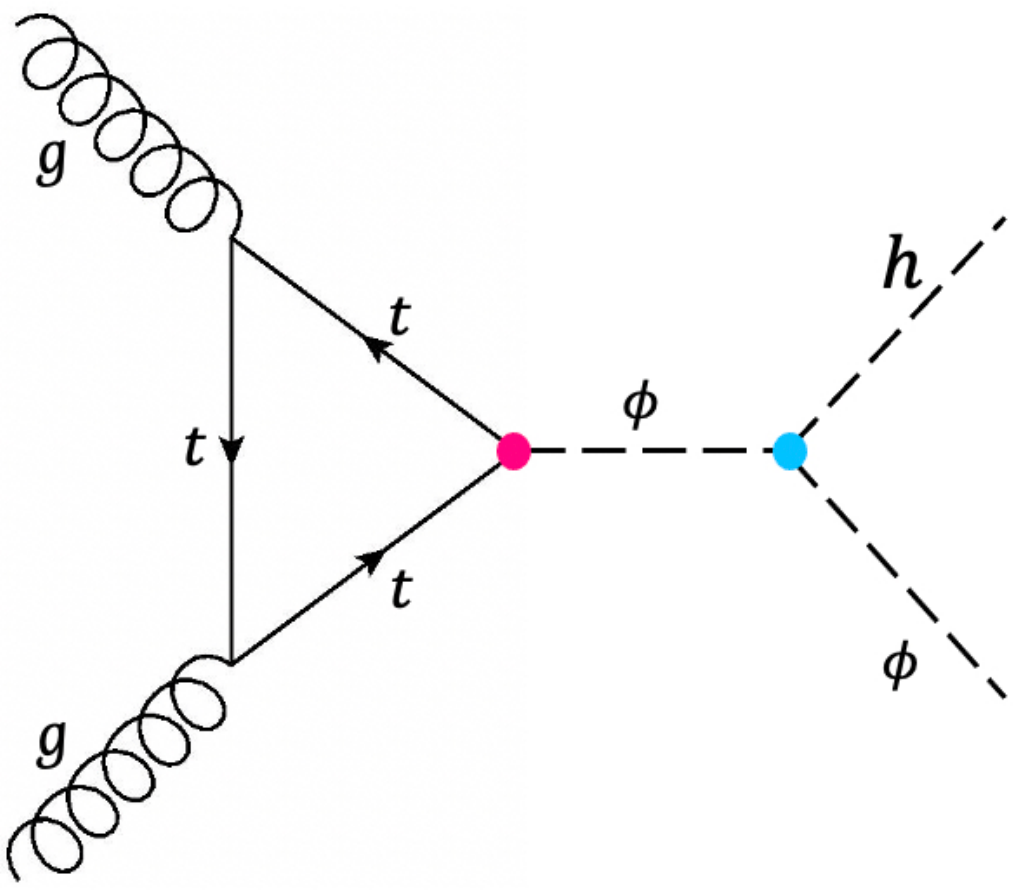}
\caption{
\label{Fig:diagram} 
The representative Feynman diagrams for $h\phi$ production in the Higgs alignment limit, where $\phi=H\,{\rm{or}}\,A$ is the additional scalar.
} 
\vspace{-.3cm}
\end{center}
\end{figure}
%%%%%%%%%%%%%%%%%%%%%%%%%%%%%%%%%%%%%%%%%%%%%%%%%

In this work, we extend this strategy to the $h\phi_l$ production.
The representative Feynman diagrams for the scalar productions are shown in Fig.~\ref{Fig:diagram}.
The left and right diagrams show the top box and triangle-induced contributions, respectively.
The magenta vertex in the diagram corresponds to the additional top-Yukawa coupling in Eq.~\eqref{Eq:Lint}, 
and the cyan one denotes the triple Higgs couplings.
The previous experimental data were not enough to probe the new physics scenarios via this production when the additional scalar boson is heavy. 
However, the expected data at the current and future LHC 
can shed light on this production as we will show.
Note that in the general 2HDM parameter space, 
for the $h\phi$ production there is another relevant diagram 
which comes from the $Z$-boson exchange Drell-Yan process. 
Although it is generated by the EW interaction, thanks to its tree-level nature, its contribution can be 
potentially much larger than the gluon-fusion process in Fig.~\ref{Fig:diagram} according to \cite{Enberg:2016ygw}.
Nevertheless, it is important to notice that such EW production is significantly suppressed in the Higgs alignment limit, while the top-loop induced productions are not suppressed.

As a demonstration,
we consider a case of $\phi_l=A$ and $\phi_h=H $. 
We note that the CP-conserving scalar potential does not allow the $hA$ production through $ gg \to h$ or $H \to hA$.
Furthermore, the Higgs alignment limit ensures that 
$ gg \to h\to hH$ process vanishes
but leaves a non-vanishing triple-Higgs couplings $\lambda_{hAA}$ and $\lambda_{hHH}$; hence  $ gg \to A\to hA$ and $ gg \to H\to hH$ processes are irreducible in the Higgs alignment limit.
The representative diagrams are shown in Fig.~\ref{Fig:diagram}.

In this study, there are four
relevant free parameters to the $h\phi_l$ production:
(1) the produced lighter scalar mass $m_{\phi_l}$, 
(2) the heavier scalar mass $m_{\phi_h}$,
(3) the triple Higgs coupling $\lambda_{hAA}$ ($\lambda_{hHH}$) for $hA$ ($hH$) production, 
and (4) the additional top-Yukawa coupling $\rho_{tt}^{\phi}$.
In our analysis, we fix $m_{\phi_l}$ and  $\rho_{tt}^{\phi}$ and 
vary $m_{\phi_h}$ and $\lambda_3$ coupling by assuming $m_{H^\pm} = m_{\phi_h}$ to avoid the experimental bound from the oblique parameter.
We numerically calculate the production cross section by {\sc\small MadGraph}5\_a{\sc\small MC}@{\sc\small NLO}~\cite{Alwall:2014hca} for a given set of $A$ and $H$ masses and $\lambda_3$.
For each benchmark parameter point, ten thousand events are generated.\footnote{
The numerical data of the production cross sections as well as the process and parameter cards for these analyses  are available in the supplementary files in \cite{data}.
}
It is noted that the matrix elements of one-loop-induced $gg \to h A$ process have been validated against an independent implementation of 2HDM model in {\sc OpenLoops2}~\cite{Buccioni:2019sur}.

\begin{figure}[t]
\begin{center}
\includegraphics[width=18.2em]{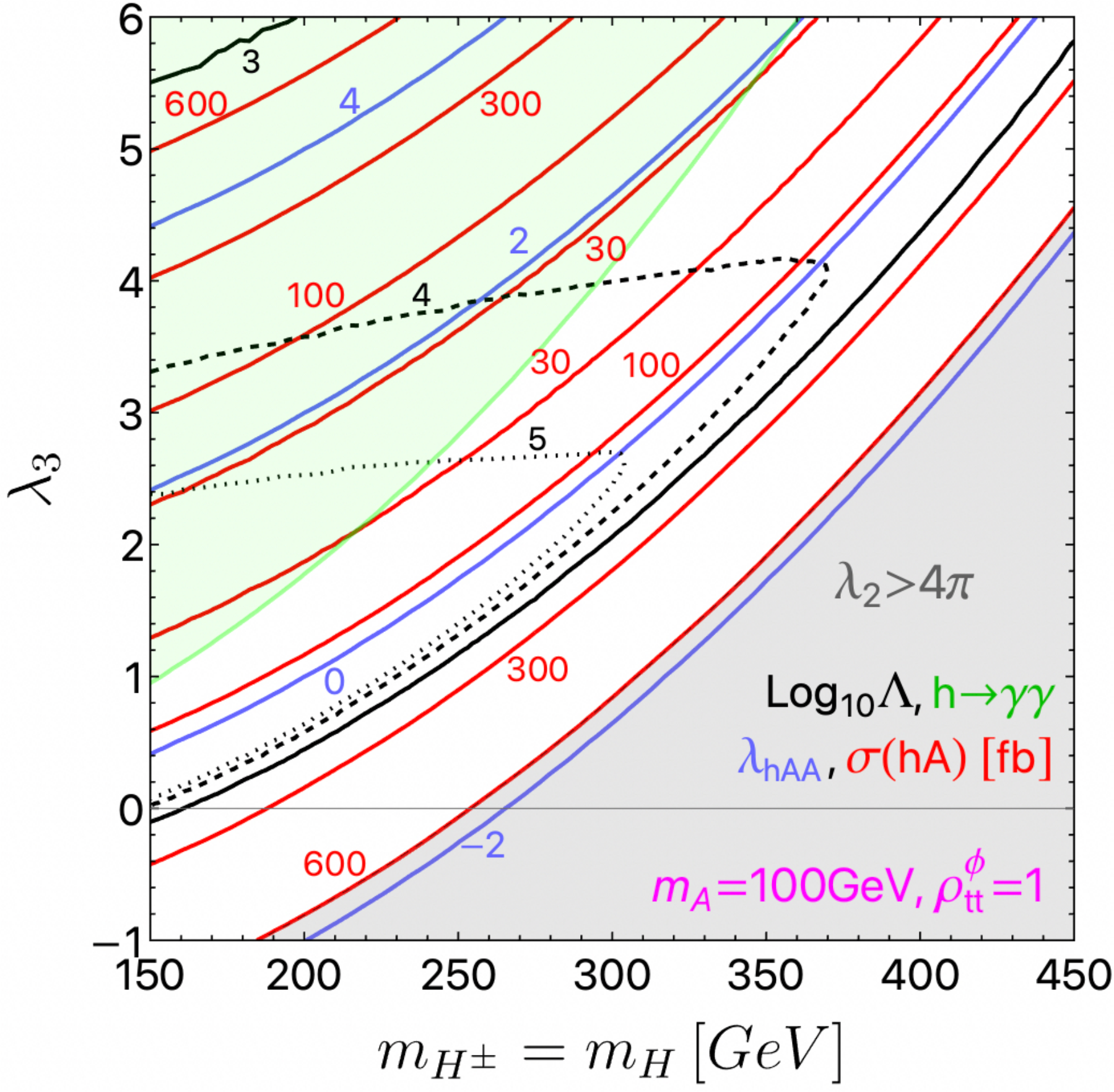}~~
\includegraphics[width=18.3em]{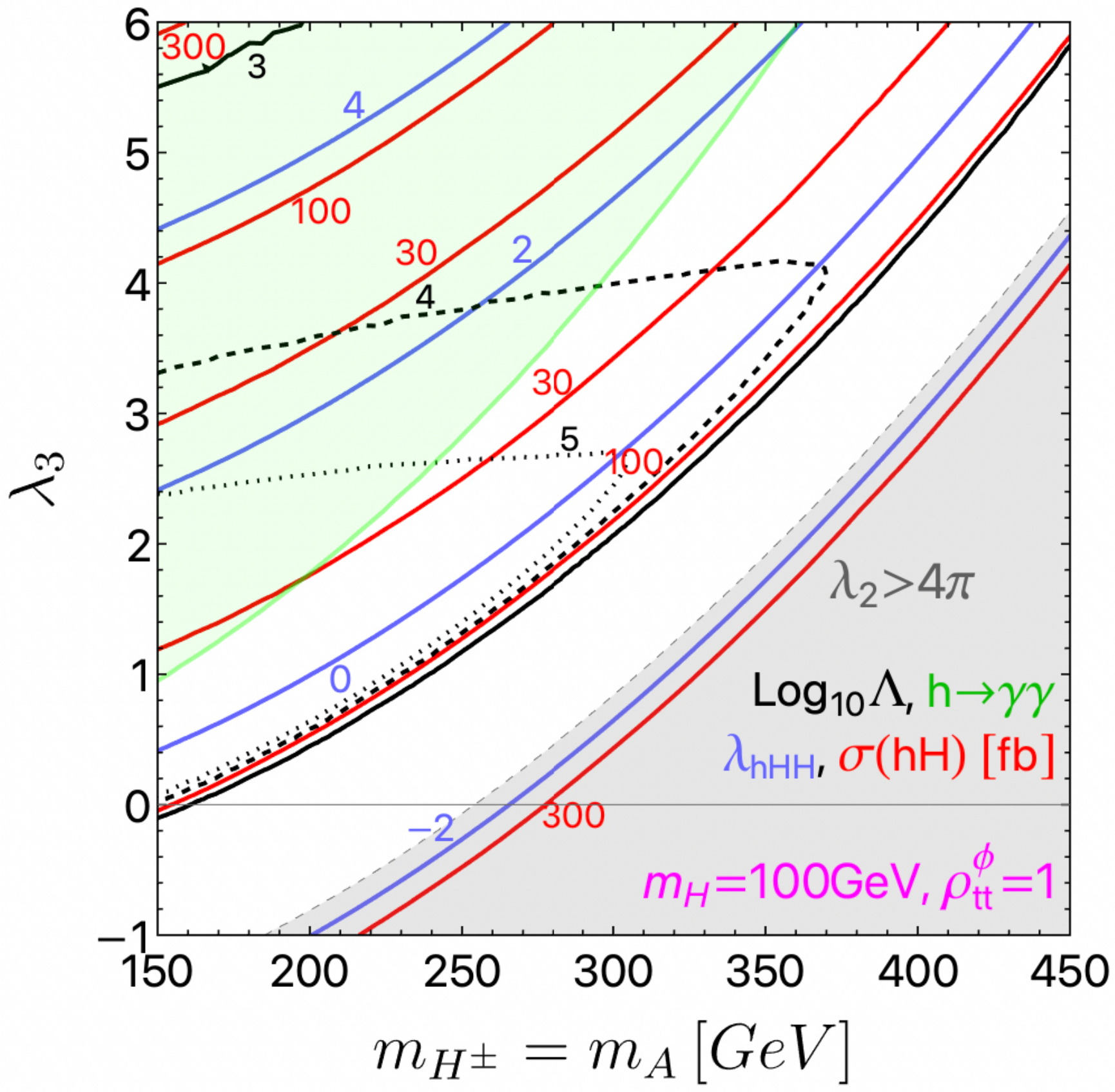}
\caption{\label{fig:Xs1} 
The $h\phi_l$ production cross section is shown in red contours on the heavier scalar mass and $\lambda_3$ plane.
The lighter scalar mass is fixed as $m_A=100\,$GeV\,(left) and $m_H=100\,$GeV\,(right).
The figures with other masses are found in Appendix \ref{sec:additionalfigure}.
Blue contours denote $\lambda_{hAA}$ and $\lambda_{hHH}$ at the additional scalar scale.
The light green region and gray regions are excluded by the $h\to\gamma\gamma$ constraint and the vacuum stability condition, respectively.
The black contours show the cutoff scale in Log$_{10}(\Lambda$/GeV).
}
\vspace{-.3cm}
\end{center}
\end{figure}

In Fig.~\ref{fig:Xs1}, by fixing $m_{\phi_l}=100$\,GeV, we show the $h \phi_l$ production cross section in the unit of fb and the $\lambda_{h\phi_l \phi_l}$ in red and blue contours, respectively.
The result for $m_{\phi_l}=50,\,150,\,200,\,250,\,300\,, 400\,$GeV can be found in Appendix~\ref{sec:additionalfigure}.
The vertical and horizontal axes correspond to $\lambda_3$ and $m_{\phi_h}$. 
We set $\rho_{tt}^\phi=1$ for simplicity since both the amplitudes corresponding to diagrams in Fig.~\ref{Fig:diagram} are linearly proportional to $\rho_{tt}^\phi$.
The cross section is proportional to $|\rho_{tt}^{\phi}|^2$, thus it is easy to rescale the cross section.
%%%%%%%%%%%%%%%%%%%%%%%%
%%%%%%%%%%%%%%%%%%%%%%%%%

The relevant constraints on $\rho_{tt}^\phi$, which depends on the scalar mass spectrum, are discussed in Sec.~\ref{sec:const_ptt}. 
Combining those constraints and production cross section in Figs.~\ref{fig:Xs1}, \ref{fig:Xs2} and \ref{fig:Xs3}, the realistic production cross section can  be obtained.
We observed that the SM-like destructive interference among the diagrams in Fig.~\ref{Fig:diagram} exists in the $hA$ as well as the $hH$ production modes. 
Especially for the $hA$ production, the additional $\gamma_5$ insertions play a similar role in the Dirac algebra of loop numerators in the box and triangle diagrams, hence the destructive interference is expected.
Our numerical analysis explicitly shows the destructive behavior in the parameter regions with $\lambda_{h \phi_l \phi_l}>0$.
It is observed that the production cross section can be $\mathcal{O}(100)$\,fb for $m_{\phi_l}=100\,$GeV with $\rho_{tt}^\phi=1$.
For $m_{\phi_l}=400\,$GeV the cross section can be $\mathcal{O}(10)$\,fb.
We see that the cancellation occurs when $\lambda_{h\phi_l\phi_l}$ is of $\mathcal{O}(1)$.

Given the similarity between $h \phi$ production and the SM $hh$ production at the leading order and the fact that higher-order QCD corrections do not introduce additional $\gamma_5$ insertions,
we further expect  that a SM-like NLO QCD $K$-factor is  around 2
for the loop-induced $gg \to h \phi$ cross sections as well.
This $K$-factor $\simeq 2$ 
is partially confirmed in Ref.~\cite{Abouabid:2021yvw} in the heavy top-mass limit, 
and it is important to confirm this expectation in the future by exact calculations which are out of the scope of this paper.

Furthermore, the Higgs precision data can constrain the parameter space.
While most of the Higgs signal strengths approach the SM Higgs predictions in the Higgs alignment limit,
the signal strength for $\gamma \gamma$ 
does not.
It is because 
the one-loop induced charged scalar contribution modifies $h\to \gamma\gamma$ rate, which has been measured with LHC Run\,2 full data \cite{ATLAS:2022tnm}.
The light green region in Fig.~\ref{fig:Xs1} is excluded at the $2\sigma$ level.
This constraint is especially relevant for the low-mass $H^\pm$ region with large coupling $\lambda_3$.
On the other hand, one could also consider the $pp \to \phi_l \to \gamma \gamma$ search.
However, the value of BR$(\phi_l \to \gamma\gamma)$ significantly depends on the Yukawa couplings $\rho_{tt}^\phi$, $\rho_{\tau \tau}^\phi$ and also $\rho_{bb}^\phi$.
So, we conclude that the constraint from the $pp \to \phi_l \to \gamma \gamma$ search is reducible in Fig.~\ref{fig:Xs1}.
Note that when one chooses the appropriate size of Yukawa couplings,  the CMS di-photon excess \cite{CMS:2018cyk} can be explained by $pp \to \phi_l \to \gamma \gamma$ \cite{Iguro:2022dok}.

If the trilinear scalar couplings are too large,
they eventually blow up 
at high energies due to the renormalization group evolution.
We consider a perturbative unitarity bound~\cite{Kanemura:1993hm,Akeroyd:2000wc,Arhrib:2000is} where the renormalization group evolution effect is considered based on Refs.~\cite{Goudelis:2013uca,Iguro:2022tmr}.
Here, we set $\lambda_7 = 0$ as a reference value. When one takes $\lambda_7 = \mathcal{O}(1)$ which is an irrelevant parameter to $h \phi_l$ production, the unitarity bound becomes more severe.
The black solid, dashed and dotted lines in Fig.~\ref{fig:Xs1} corresponds to the cutoff scale of $\Lambda =10^3, 10^4$ and $10^5 \,{\rm{ GeV}}$, respectively.
If one requires the model to be perturbative up to $10$\,TeV, then the region outside the black dashed line will be excluded. 

Another theoretical constraint on the quartic couplings comes from the vacuum stability of the scalar potential.
In this paper, we impose the tree-level bounded-from-below condition for the scalar potential \cite{Maniatis:2006fs,Ferreira:2009jb},
\begin{align}
\lambda_1, \lambda_2 > 0\,, \quad 
\sqrt{\lambda_1\lambda_2}+\lambda_3 > 0\,,\quad
\sqrt{\lambda_1\lambda_2} +\lambda_3+\lambda_4-|\lambda_5| >0\,. 
\label{eq:BFB}
\end{align}
In order to produce a large mass difference between $\phi_l $ and $\phi_h$, $\lambda_4$ must be largely negative value at the EW scale. 
Therefore, to satisfy the last condition in Eq.~\eqref{eq:BFB}, $\lambda_2$ must be largely positive, though it does not change the phenomenology discussed in this paper.
The gray region in Fig.~\ref{fig:Xs1} is excluded  since there $\lambda_2$ becomes too large ($\lambda_2 > 4\pi$) to respect the bounded-from-bellow condition \cite{Deshpande:1977rw}. 
We see that various constraints are very complementary on this plane.

%%%%%%%%%%%%%%%%%%%%%%%%%%%%%%%
\subsection{Constraints on \texorpdfstring{$\rho_{tt}^\phi$}{rho phi tt}}
\label{sec:const_ptt}
%%%%%%%%%%%%%%%%%%%%%%%%%%%%%%%
In this section,
we discuss the relevant flavor and collider constraints on the  top-Yukawa couplings.
First of all, we summarize the constraints from the direct searches at the LHC. 
The $tb$ resonance search is relevant for the additional charged scalar $H^\pm$, and the $t\ov{t}$ resonance search is relevant for the extra neutral scalars $H$ and $A$ when $\rho_{tt}^\phi$ is dominant among the interactions. 
On the other hand, 
since a bound from 
the $\tau\ov{\tau}$ resonance search strongly depends on whether the scalar has other decay modes or not,
we consider only the case where $\rho_{tt}^\phi$ is dominant compared to other couplings for simplicity.

\begin{figure}[t]
\begin{center}
\includegraphics[width=20em]{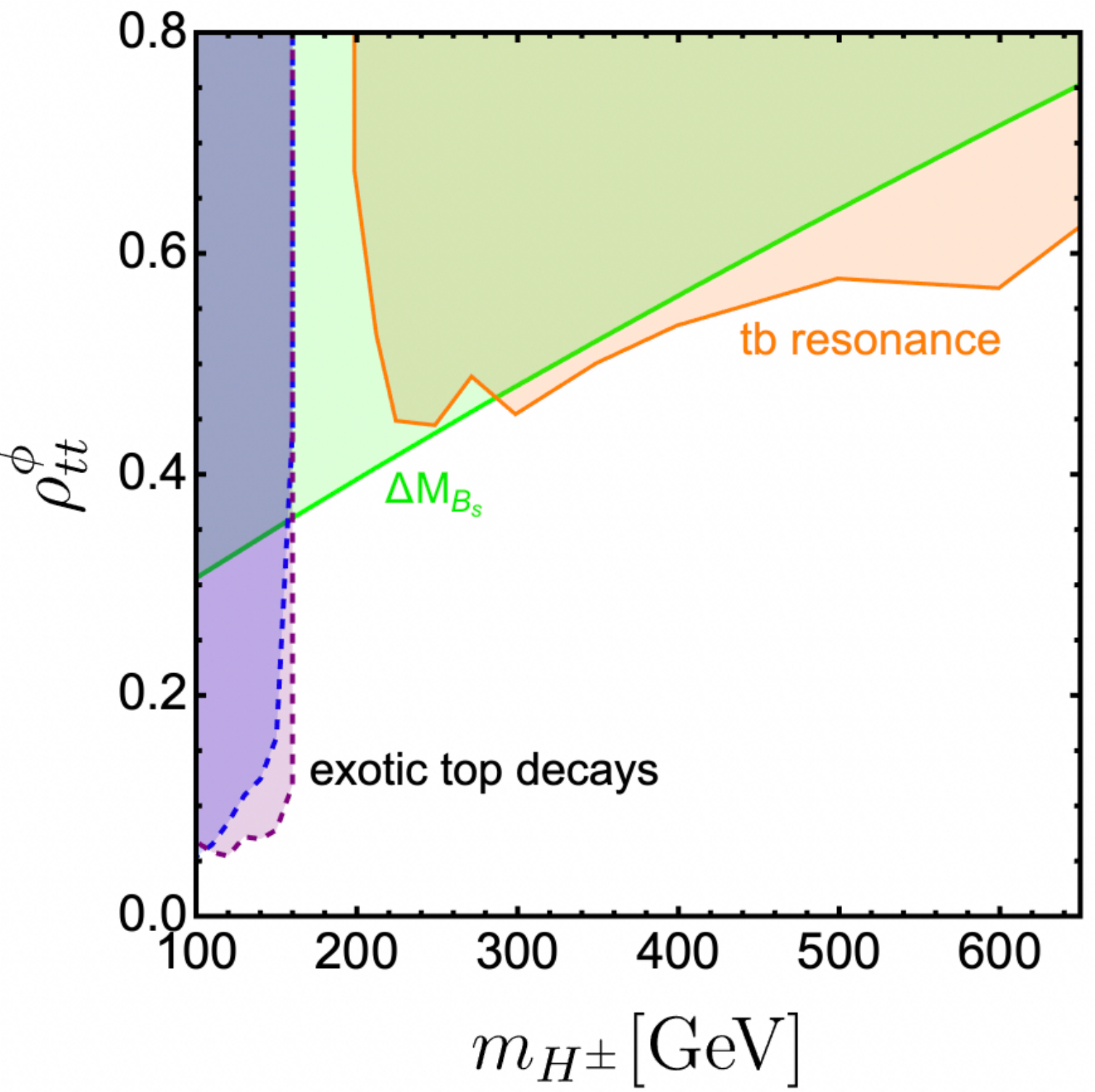}
\caption{
The upper limit on $\rho_{tt}^\phi$ as a function of $m_{H^\pm}$ at 95$\%$ CL.
Orange region is excluded by the $tb$ resonance search.
Purple and blue regions surrounded by dotted lines could be disfavored by the $t\to b H^\pm\to b \tau\nu$ and $t\to b H^\pm\to b b c$ decays, respectively.
The light green region is excluded by the $B_s$-meson mixing bound.
}
\label{Fig:ULptt}
\vspace{-.3cm}
\end{center}
\end{figure}

In Fig.~\ref{Fig:ULptt},  
we derive the upper limit on $\rho_{tt}^\phi$ as a function of $m_{H^\pm}\,(=m_{\phi_h})$.
The orange region is excluded by the
$pp\to tbH^\pm\to  \bar{t}b t\bar{b}$ process based on the LHC Run\,2 full data~\cite{ATLAS:2021upq} assuming $\text{BR}(H^\pm\to tb) =1$.
It is noted that the experimental data is available in the region of  $m_{H^\pm}\ge200\,$GeV.
In fact,
there is also a decay mode where the scalar decays into gauge bosons
$H^\pm\to W^\pm \phi_l$, and the $\bar{t}b$ branching fraction is diluted for $m_{H^\pm}-m_{\phi_l} > m_W$.
For instance, 
we obtain $\text{BR}(H^\pm\to tb)=0.6$ for $\rho_{tt}^\phi=1$ with $m_{H^\pm}=350$\,GeV and $m_{\phi_l}=100$\,GeV.
Notice that the
dilution of the bound significantly depends on $m_{\phi_l}$.

Another collider constraint for a light $H^\pm$ scenario comes from the exotic top decay which is available in the range of $m_{H^\pm}\le 160\,$GeV.
If the couplings other than $\rho_{tt}^\phi$ are negligible for $m_{H^\pm}\le m_t$, then $H^\pm\to cb$ will be the dominant decay mode via $\rho^d_{bb}$ or $\rho^u_{tc}$ interactions in Eq.~\eqref{eq:Lintgenenral}. 
Both ATLAS and CMS collaborations have searched for $H^\pm\to cb$ channel via 
a top quark decay \cite{CMS:2018dzl,ATLAS:2021zyv},
and they set the upper limit on the product of $\text{BR}(t\to b H^\pm)\times \text{BR}(H^\pm \to bc)$
as a function of $m_{H^\pm}$.
Assuming $\text{BR}(H^\pm\to cb)=1$, we can set the upper limit on $\rho_{tt}^\phi$ which is shown by the blue region in Fig.~\ref{Fig:ULptt}.
Similarly, if $H^\pm\to\tau\nu$ is the dominant decay mode via scalar-tau Yukawa interaction, we can set the upper limit on $\rho_{tt}^\phi$ which is shown by the purple region in Fig.~\ref{Fig:ULptt} \cite{ATLAS:2018gfm}.
Furthermore, the similar upper limit can be obtained from  $H^\pm\to cs$ via $\rho^u_{cc}$ and $\rho^d_{ss}$~\cite{CMS:2020osd}.
It is noted that $t\ov{t}$ resonance and four-top searches give a weaker constraint \cite{CMS:2018rkg}, and thus they are omitted.
%%%%%%%%%%%%%%%%%%%%%%%%
%%%%%%%%%%%%%%%%%%%%%%%%%

The large $\rho_{tt}^\phi$ also modifies the $B_{(s)}$-meson mixing 
$\Delta M_{B_{(s)}}$ at one-loop level.
We use the analytic formula for the $H^\pm$ box contribution ($H^\pm$-$W^\mp$ and $H^\pm$-$H^\mp$ boxes) from Refs.~\cite{Iguro:2017ysu,Iguro:2018qzf} and adopt the latest bound~\cite{DiLuzio:2019jyq}.
The renormalization group running correction from $\mu=m_{H^\pm}$ to $\mu=m_{W}$ is taken into account.
Notice that other flavor constraints, for example, the ones from $b\to s\gamma$ and semileptonic decays of kaon, are confirmed to be less stringent \cite{Iguro:2019zlc}.
In Fig.~\ref{Fig:ULptt}, the light green region is excluded by the $B_s$ mixing constraint.
Since the $B_s$ mixing constraint does not depend on $m_{\phi_l}$ and the decay modes, 
the bound is the most conservative in Fig.~\ref{Fig:ULptt}.
It is observed that the mass gap exists for the LHC bounds around $m_{H^\pm}\sim m_t$ due to the experimental difficulty.
However, the $B_s$ mixing constraint is complementary and thus fills the gap.

It is observed that for example, $\rho_{tt}^\phi\simeq 0.4$ is allowed for the $m_{H^\pm} \gtrsim 200$\,GeV scenario.
Combining these constraints with Figs.~\ref{fig:Xs1}, \ref{fig:Xs2} and \ref{fig:Xs3}, one can read the maximal production cross section of the $h \phi_l$ channels as we will show an example in the next section.  

It is worth mentioning that 
in addition to the $\rho^\phi_{tt}$ bound, 
a severe constraint on $\lambda_{h\phi_l\phi_l}$ appears 
for $m_{\phi_l}\le m_h/2$ in Fig.~\ref{fig:Xs2}.
In this parameter region, the
$h\to \phi_l\phi_l$ decay is kinematically open  and modifies the Higgs total width.
The partial decay width is given as 
\begin{align}
    \Gamma(h\to \phi_l\phi_l)=\frac{\lambda_{h\phi_l\phi_l}^2 v^2}{32\pi m_h}\sqrt{1-\frac{4 m_{\phi_l}^2}{m_h^2}}\,.
\end{align}
The current bound \cite{ATLAS:2020cjb,CMS:2022qva} corresponds to $\lambda_{h\phi_l\phi_l}\le {\mathcal{O}}(10^{-2})$ for $m_{\phi_l}\le m_h/2$.

%%%%%%%%%%%%%%%%%%%%%%%%%%%%%%%
\subsection{Comment on solutions of di-tau excess and muon \texorpdfstring{$g-2$}{g-2}}
\label{sec:impact}
%%%%%%%%%%%%%%%%%%%%%%%%%%%%%%%
In this subsection, we briefly discuss the phenomenological impacts on the di-tau excess and the muon $g-2$ anomaly explanations in the 2HDMs.

Since the 100\,GeV $\tau\bar{\tau}$ excess requires an additional light scalar  decaying into di-tau final states, we introduce the following interaction,
\begin{align}
-\mathcal{L}_{\rm Yukawa}^\tau =
\frac{\rho_{\tau\tau}^\phi}{\sqrt{2}}\,\bar{\tau} H \tau+
i \frac{\rho_{\tau\tau}^\phi}{\sqrt{2}}\,\bar{\tau} A \gamma_5 \tau\,.
\label{Eq:LYtau}
\end{align}
It is shown in Ref.~\cite{Iguro:2022dok}  that the a simultaneous explanation of the di-tau \cite{CMS:2022goy} and di-photon  excesses \cite{CMS:2018cyk}  favors
$m_{\phi_l} = m_A\simeq95\text{--}100$\,GeV with $\rho_{tt}^\phi\simeq 0.4$ and   $\rho_{\tau\tau}^\phi \simeq 3\times 10^{-3}$, which corresponds to $\text{BR}(A\to\tau\ov\tau)\simeq 0.3$.
As seen from Fig.~\ref{Fig:ULptt} that $m_{H^\pm}\le$ 200\,GeV with $\rho_{tt}^\phi\simeq 0.4$ can not satisfy the  $\Delta M_{B_s}$ bound.
Furthermore, we found that $m_H\le2m_t$ is excluded by the di-tau search \cite{CMS:2022rbd}.\footnote{It is noted that $m_{H^\pm}\le m_t+m_b$ is also disfavored by the $\tau\nu$ resonance search associated with the top and bottom quarks \cite{ATLAS:2018gfm}.}
For $m_H\ge2m_t$,  $\text{BR}(H\to\tau\ov\tau)$ is  significantly diluted and the bound becomes  weak, thus the explanation is possible.

On the other hand, the current $95\%$ CL upper limits on the signal strengths of the di-Higgs production in $b\ov{b}\tau\ov{\tau}$ decay mode is given as $\mu(hh\to b\ov{b}\tau\ov{\tau})\le3.3$ \cite{CMS:2022hgz}. 
The NNLO SM prediction is given as $\sigma_{\rm ggF}^{hh}=31.05^{+6\%}_{-23\%}\pm3\%$\,fb with $m_h=125\,$GeV \cite{Grazzini:2018bsd,Dreyer:2018qbw}.
It is noted that the experimental upper limit is a consequence of the combination of gluon fusion (ggF) and vector boson fusion (VBF).
Although the VBF cross section is small,  $\sigma_{\rm VBF}^{hh}=1.73^{+0.03\%}_{-0.04\%}\pm2.1\%$\,fb at NNLO \cite{Dreyer:2018qbw}, the unique VBF event topology provides a useful handle to identify signal events and sensitivity.
While it is nontrivial to separate the impacts of ggF and VBF determination, 
the ggF mainly determines the upper limit \cite{CMS:2022hgz}.
With the SM predictions 
$\text{BR}(h\to b\ov{b})=58.1\%$ and $\text{BR}(h\to\tau\ov{\tau})=6.3\%$, we can estimate a rough sensitivity of $hA$ production at the LHC
by only considering the ggF mode,
and the current LHC data 
reaches $7.5\,\rm{fb}$ for the di-Higgs production with $b\ov{b}\tau\ov{\tau}$ final state. 
%$3.3\times 31.05\,\rm{fb}\times 2\times 58\%\times6.3\%= 3.3\times 2.3 =7.5\,\rm{fb}$.

From Fig.~\ref{fig:Xs1}, we can see that $\sigma(pp \to hA)$ is as large as $200$\,fb for $\rho_{tt}^\phi=1$ when the cutoff scale $\Lambda > 10$\,TeV is imposed.
For $\rho_{tt}^\phi=0.4$, the production cross section is calculated as  $\sigma(pp \to hA)=200\times0.4^2\simeq30$\,fb.
Therefore the explanation of $\tau \ov{\tau}$ excess with $m_A\sim100$GeV, $\rho_{tt}^\phi=0.4$,  $\rho_{\tau\tau}^\phi=3\times 10^{-3}$
predicts 
$\sigma(pp\to hA)\times \text{BR}(h\to b\ov{b})\times \text{BR}(A\to\tau\ov{\tau})\simeq 5.6$\,fb, which is larger than the SM di-Higgs prediction ($2.3$\,fb).
%$\sigma(hA\to b\ov{b}\tau\ov{\tau})\simeq 200\,\rm{fb}\times 0.4^2\times 58\%\times 30\%\simeq 5.6$\,fb which is larger than SM prediction.
If the $K$-factor is about 2 for $\sigma(pp \to hA)$, then the model prediction exceeds the current exclusion for the $hh$ channel.
As shown in Ref.~\cite{Iguro:2022dok},
unlike the CP-even interpretation, 
it is challenging to test the CP-odd solution for the $\tau\bar{\tau}$ excess in $t\bar{t}+\tau\bar{\tau}$ final state,
whereas
the $hA$ production (decaying into $b\ov{b}\tau\ov{\tau}$) allows us probe this solution.
In this context, more dedicated sensitivity studies are necessary from both experimental and theoretical sides.
It is noted that the single $H$ production process can be relevant for the model parameter space.
The CMS collaboration probed $pp\to H\to AZ\to \tau\bar\tau +l\bar{l}$ process with the Run\,1 data and the $\mathcal{O}(10)$\,fb upper limit on $\sigma(pp\to H\to AZ\to \tau\bar\tau +l\bar{l})$ has been set.\footnote{The ATLAS and CMS collaborations focused on only $b\bar{b}+l\bar{l}$ final state in  the Run\,2 data \cite{ATLAS:2018oht,CMS:2019ogx}.}
We calculated the signal cross section based on the {\sc\small{SusHi}}
\cite{Harlander:2012pb,Harlander:2016hcx} and confirmed that the parameter which explains $\tau\tau$ and di-photon excesses simultaneously predicts the smaller signal cross section by roughly a factor of $1/2$.
The LHC Run\,2 full result would give conclusive evidence.

It is known that a light pseudo scalar is necessary for an explanation of the muon 
$g-2$ anomaly in the type-X and flavor-aligned (FA-) 2HDMs \cite{Ferreira:2021gke}, where the dominant contribution comes from  the two-loop Barr-Zee diagram with tau loop.
In the type-X 2HDM, 
the top-Yukawa coupling with the heavy scalar is $\mathcal{O}(0.01)$ for the excess favored region.
Therefore, the di-Higgs production cross-section is too small to be observed.
We also found that even in the FA-2HDM, the top-Yukawa coupling can not be large due to the constraint from $B_s\to \mu \bar{\mu}$ measurement that is enhanced with a light scalar exchange diagram \cite{Li:2014fea}.
Furthermore, it would not be easy
to reconstruct the additional light scalar in $\tau\bar{\tau}$ final state at the LHC due to several neutrinos. 
Therefore, we conclude that the di-Higgs production for the solution of the 
muon $g-2$ anomaly is less relevant.

%%%%%%%%%%%%%%%%%%%%%%%%%%%%%%%
\section{Conclusion}
\label{sec:conclusion}
%%%%%%%%%%%%%%%%%%%%%%%%%%%%%%%
The Higgs field plays a very important role in giving masses to the SM particles. 
However, there are still open questions, for example, the origins of the negative mass term and also the hierarchy of the fermions. There would be new physics to solve the questions.
On the other hand, it is known that the extended Higgs models can be the key to explaining 
the origins of the neutrino masses, EW baryogenesis, dark matter, and EW vacuum stability.
Therefore it is often believed that the study of the Higgs sector is the way to access physics beyond the SM.
The 2HDM is one of the simplest extensions of the SM that frequently appears in new physics scenarios.
It is known that this model can explain the excesses in the $\tau\bar\tau$ and $\gamma\gamma$ resonances data reported by the CMS collaboration 
as well as the muon $g-2$ anomaly.
Those discrepancies require a light neutral scalar which will be within the reach of the LHC.
This kind of light scalar is also well motivated by a successful strong first-order EW phase transition and the baryon asymmetry of the Universe.

In order to chase a realistic and still-allowed 2HDM that can resolve the aforementioned puzzles, 
in this paper, we investigated the $h\phi$ production where $\phi$ is the additional neutral scalar ($\phi=H,\,A$). 
This process will be interesting in the HL-LHC era. 
We calculated the cross section of $h \phi$ production from the loop-induced gluon fusion channel at the leading order.
We took into account 
the theoretical bounds from the perturbative unitarity and vacuum stability conditions, and the experimental bounds from the Higgs and flavor precision measurements as well as the direct search at the LHC.
It is found that $h\phi$ production cross section could be as large as $\mathcal{O}(30)$\,fb if the additional scalar mass is around 100\,GeV.
One should note that $\phi\phi$ production (would decay into $4\tau$,  which is harder to probe than 2$b$2$\tau$ from $h\phi_l$ production) can also be a relevant search channel. 
Although the top-loop box contribution is suppressed by an additional factor of $\rho^\phi_{tt}$, $gg \to h \to \phi \phi$ and $gg \to \phi \to \phi \phi$ contributions are not, and the latter depends on the value of $\lambda_7$ and also the resonance enhancement is possible.

Furthermore, motivated by the low mass $\tau\bar{\tau}$ and $\gamma\gamma$ resonant excesses reported by the CMS collaboration,
we investigated the impact of the $h\phi$ production.
It was found that the combined explanation of the excesses predicts $\sigma(pp\to hA\to b\bar{b}\tau\bar\tau)\simeq 6\,$fb at the QCD leading order. 
Very interestingly,
this leading-order cross section is already larger than the SM Higgs-pair production decaying into $b\ov{b}\tau\ov{\tau}$ 
by a factor of more than two.
Therefore, 
this mode provides a unique window to probe the possible explanation of the excesses.
Lastly, it is clear that more dedicated precise calculations and experimental simulations to evaluate the realistic sensitivity are further needed.

%%%%%%%%%%%%%%%%%%%%%%%%%%%%%
\section*{Acknowledgements}
%%%%%%%%%%%%%%%%%%%%%%%%%%%%%
The authors would like to thank Sven Heinemeyer, Jonas Lindert, Ulrich Nierste, Masanori Tanaka, and Kei Yagyu for fruitful comments and valuable discussion.
%---------------------------------------------------------------------------
S.\,I. and T.\,K. thank the workshop ``Physics in LHC and Beyond'', where a part of discussion took place. 
%---------------------------------------------------------------------------
We also appreciate TTP KIT for the massive computational support,
especially we thank Martin Lang, Fabian Lange and Kay Sch{\"o}nwald for the computational help.
%---------------------------------------------------------------------------
S.\,I. and H.\,Z. are supported by the Deutsche Forschungsgemeinschaft (DFG, German Research Foundation) under grant 396021762-TRR\,257.
%%%
T.\,K.~is supported by the Grant-in-Aid for Early-Career Scientists from the Ministry of Education, Culture, Sports, Science, and Technology (MEXT), Japan, No.\,19K14706.
%%%
The work of Y.\,O.~is supported by Grant-in-Aid for Scientific research from the MEXT, Japan, No.\,19K03867.
%%%
This work is also supported by 
the Japan Society for the Promotion of Science (JSPS)  Core-to-Core Program, 
No.\,JPJSCCA20200002.
%---------------------------------------------------------------------------
%%%%%%%%%%%%%%%%%%%%%%%%%%%%%%%%%%%%%%%%%%%%%%%%%%%%%%%

\appendix

%%%%%%%%%%%%%%%%%%%%%%%%%%%%%%%
\section{Additional figures}
\label{sec:additionalfigure}
\begin{figure}[p]
\begin{center}
\includegraphics[width=16.5em]{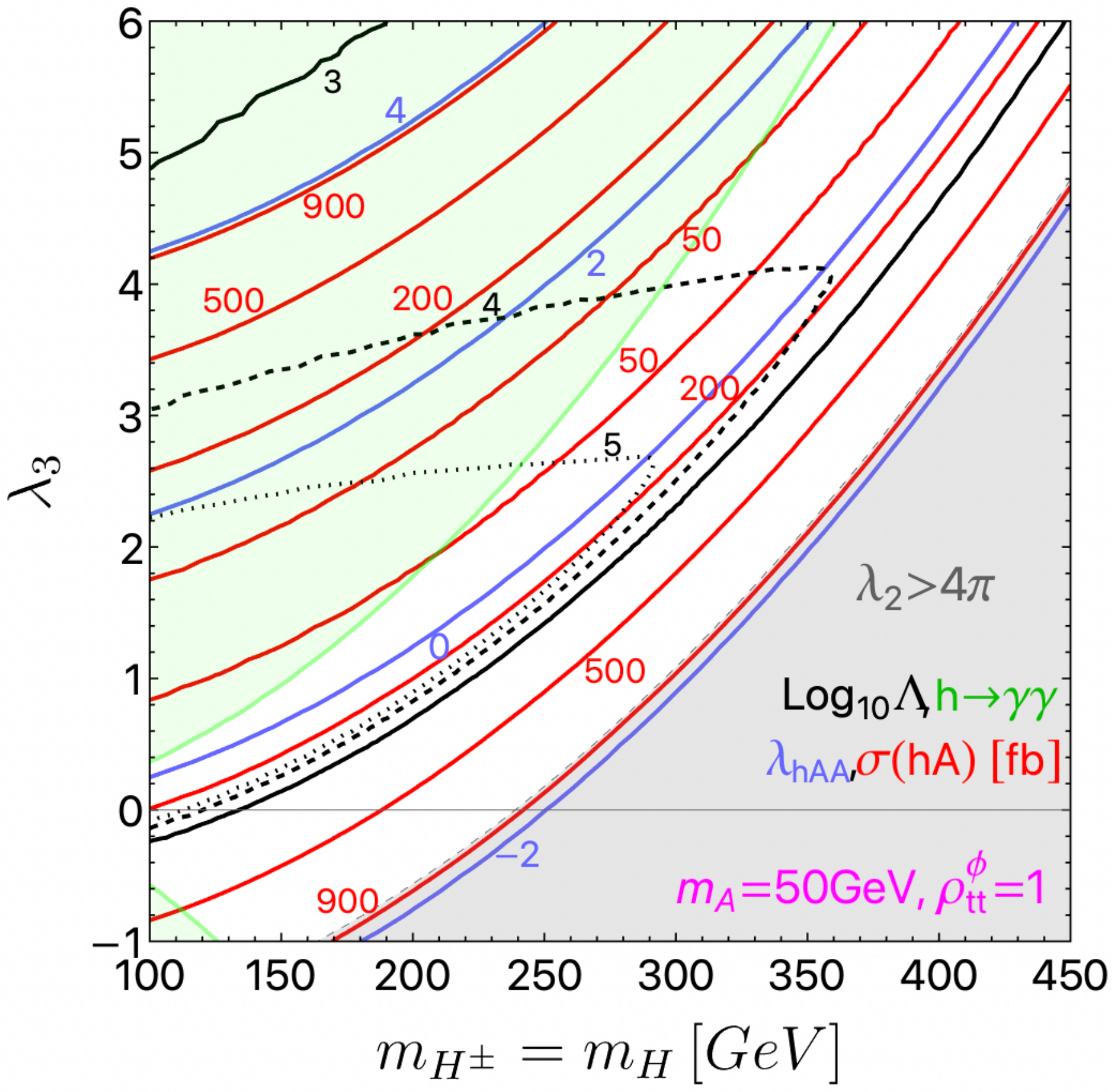}~~
\includegraphics[width=16.5em]{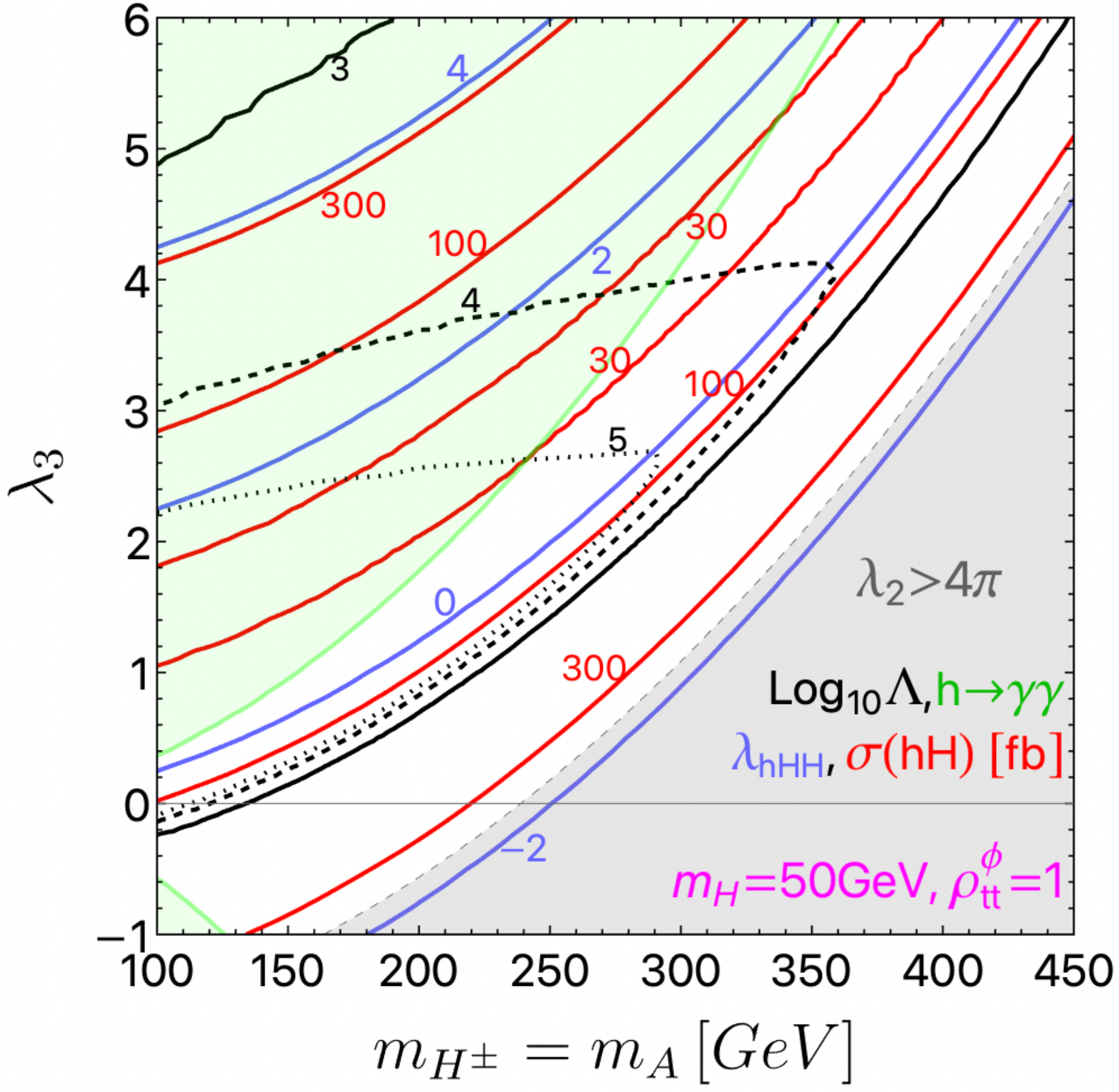}\\
\!\!\!\!\!\!
\includegraphics[width=16.em]{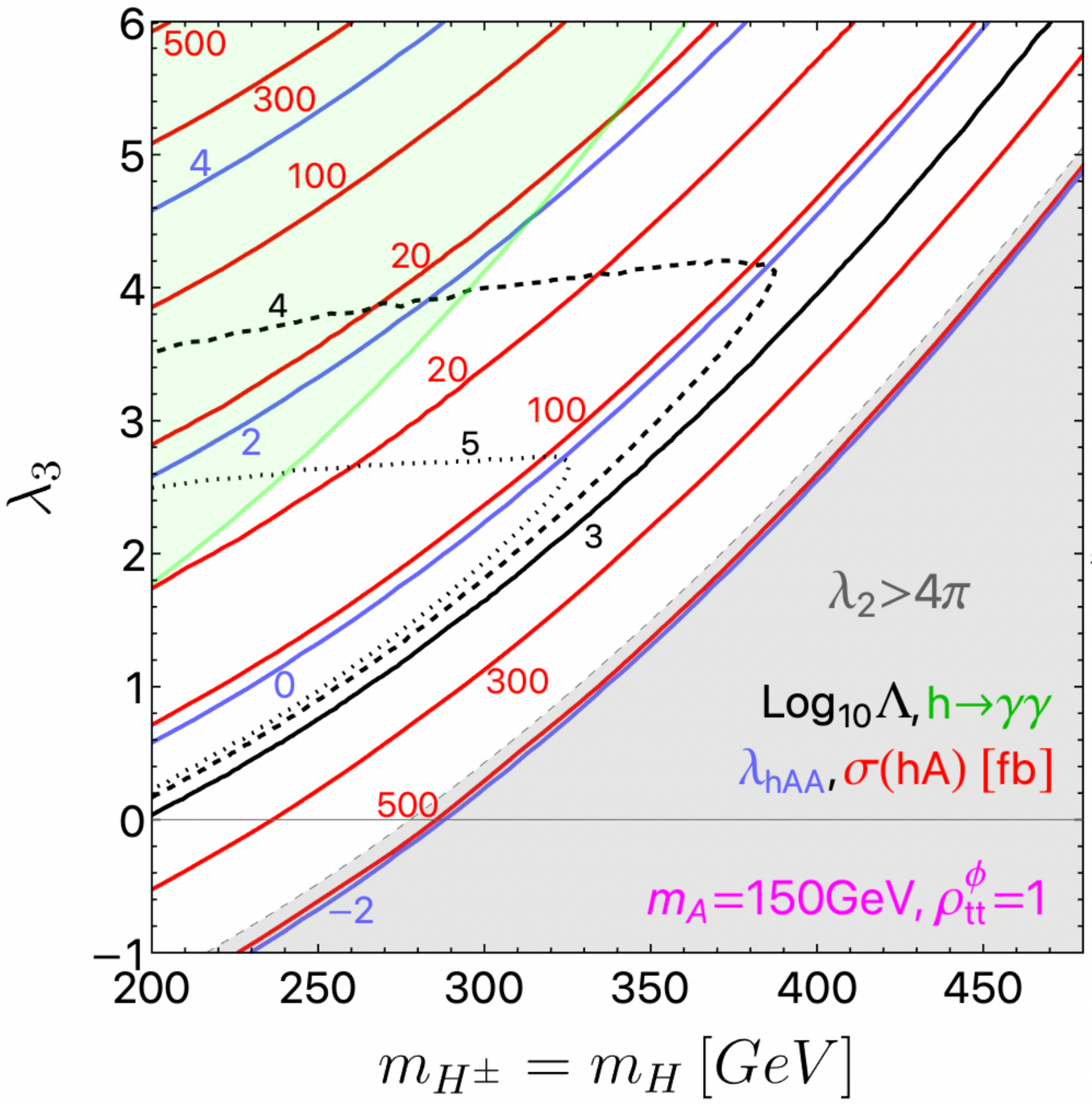}~~~~
\includegraphics[width=16.em]{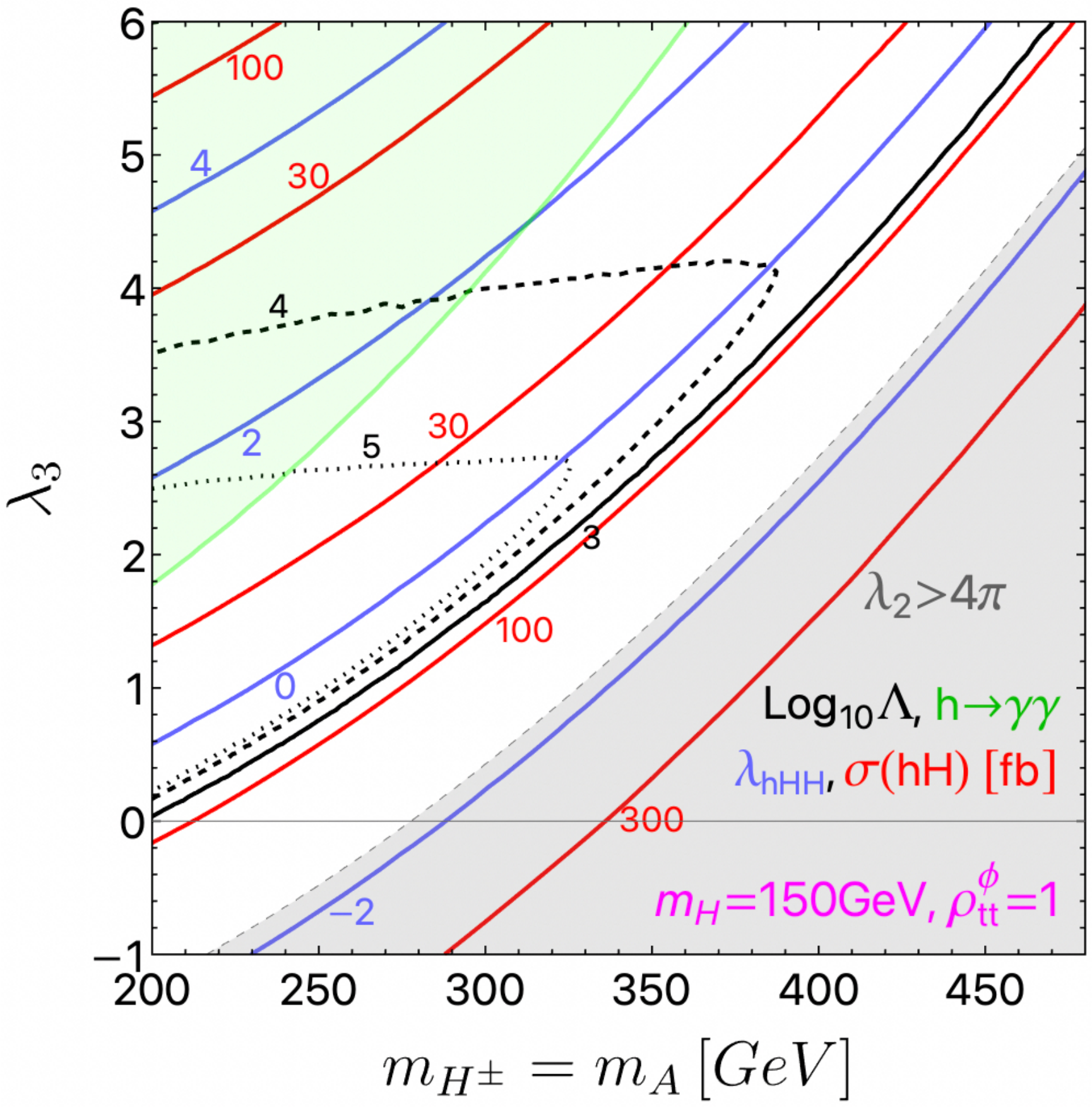}\\
\includegraphics[width=16.5em]{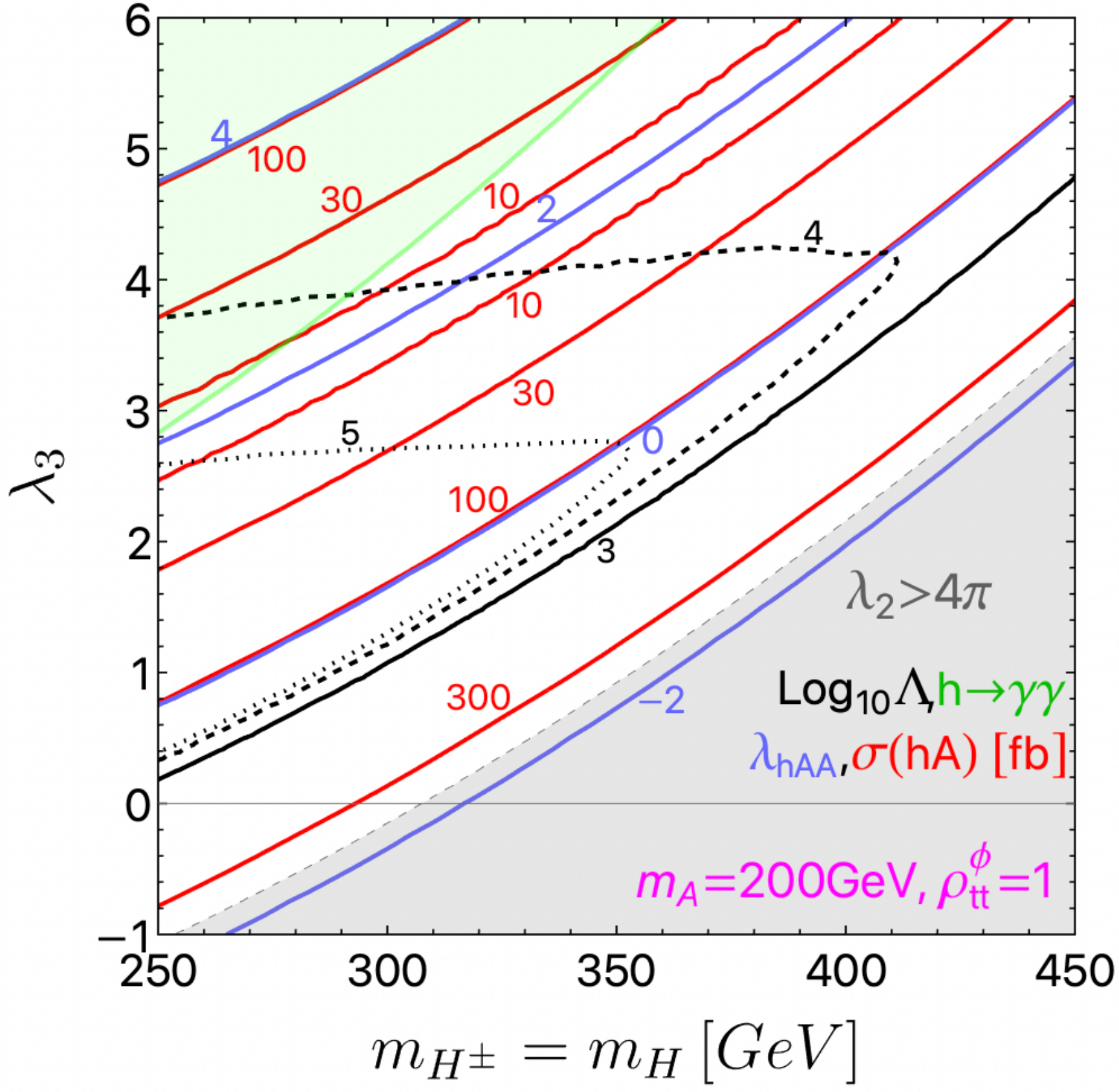}~~
\includegraphics[width=16.5em]{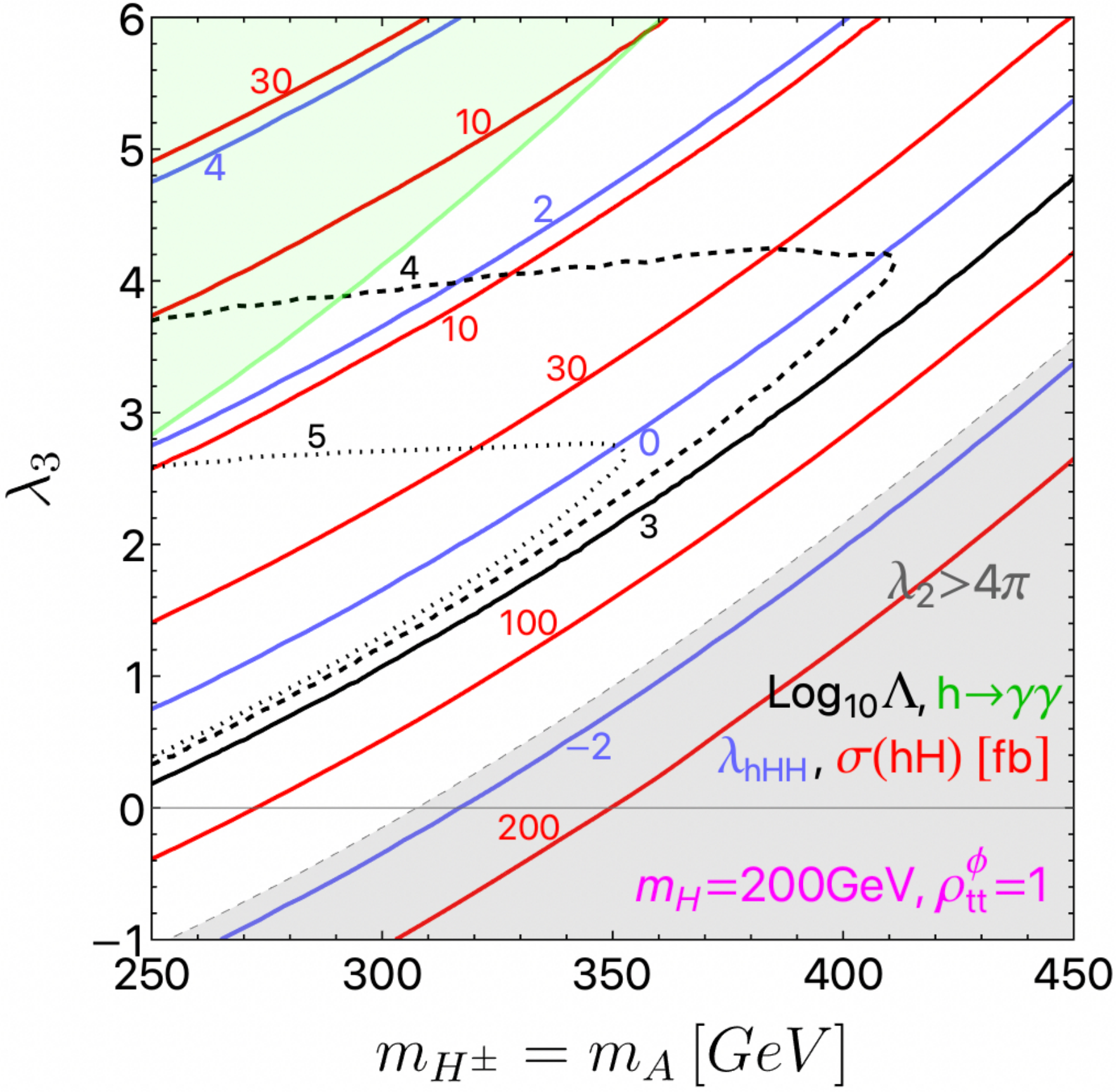}
\caption{\label{fig:Xs2}
The $h\phi_l$ production cross section is shown on the heavier scalar mass vs.\,$\lambda_3$ plane.
The left and right figures are for $hA$ and $hH$ productions, respectively.
The lighter scalar mass is fixed to be 50\,GeV\,(upper), 150\,GeV\,(middle), 200\,GeV\,(bottom).
See, the caption of Fig.~\ref{fig:Xs1} for the description of the constraints and other explanations.
}
\end{center}
\end{figure}

\begin{figure}[p]
\begin{center}
\includegraphics[width=16.5em]{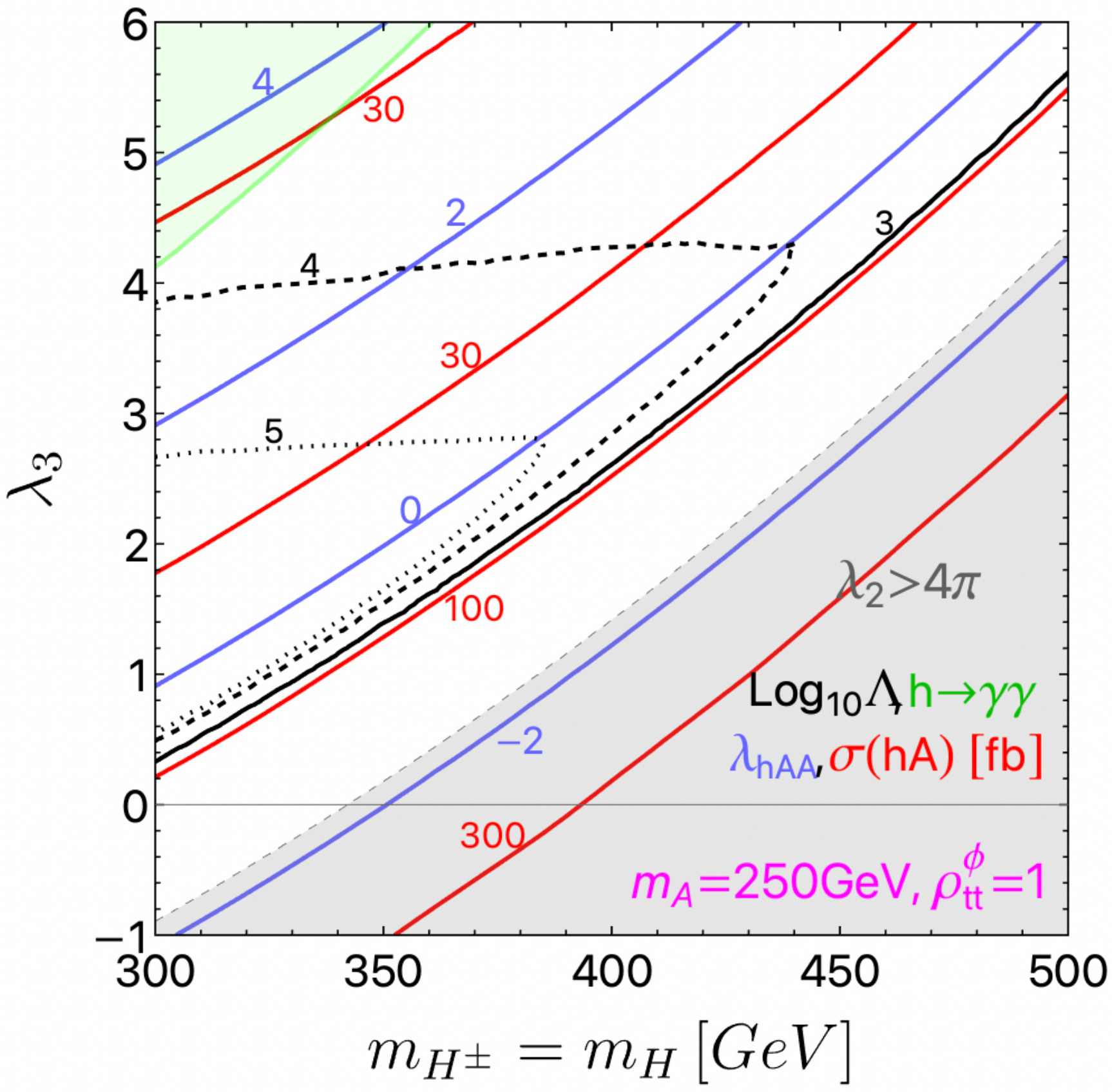}~~
\includegraphics[width=16.5em]{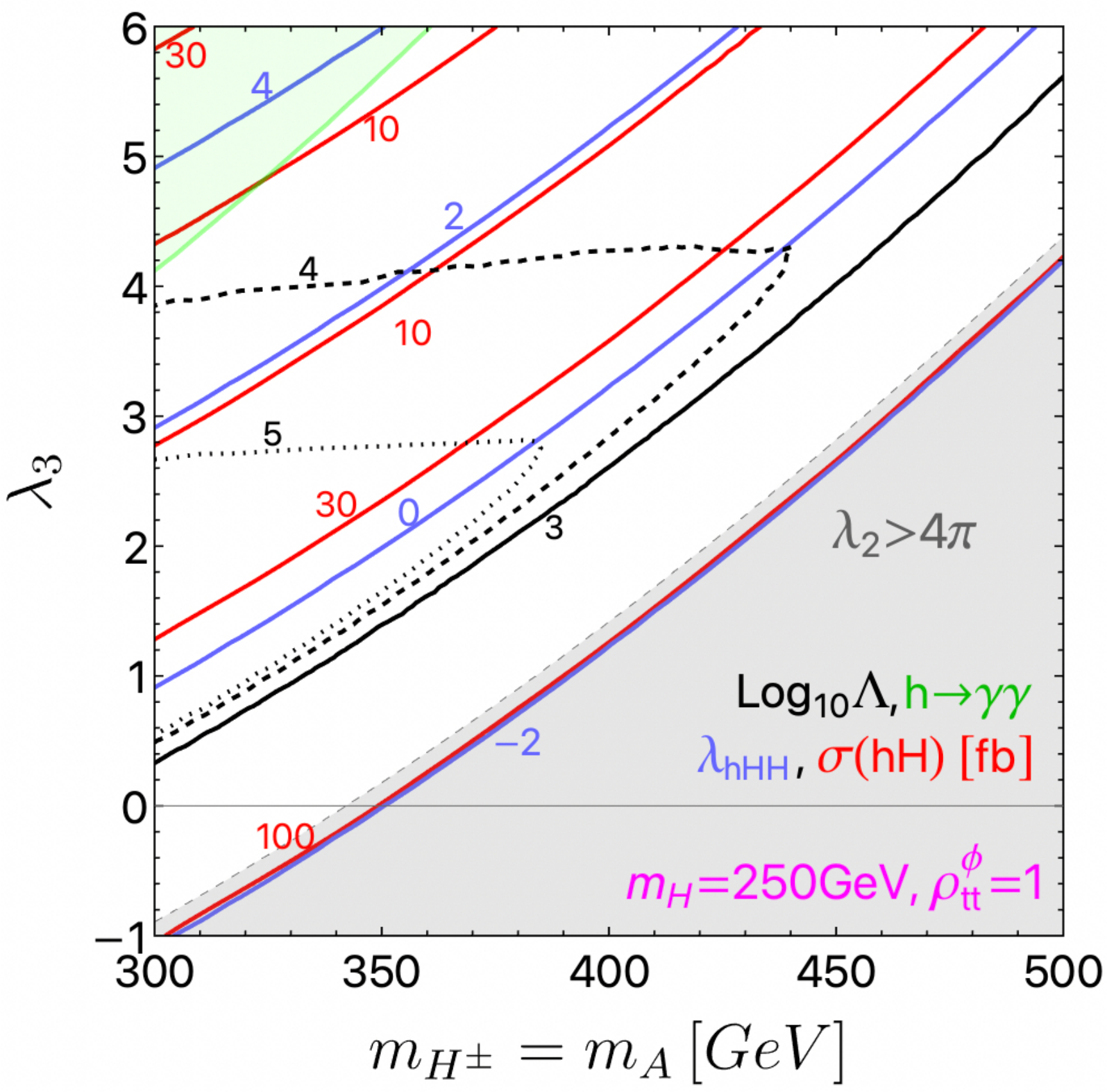}\\
\includegraphics[width=16.5em]{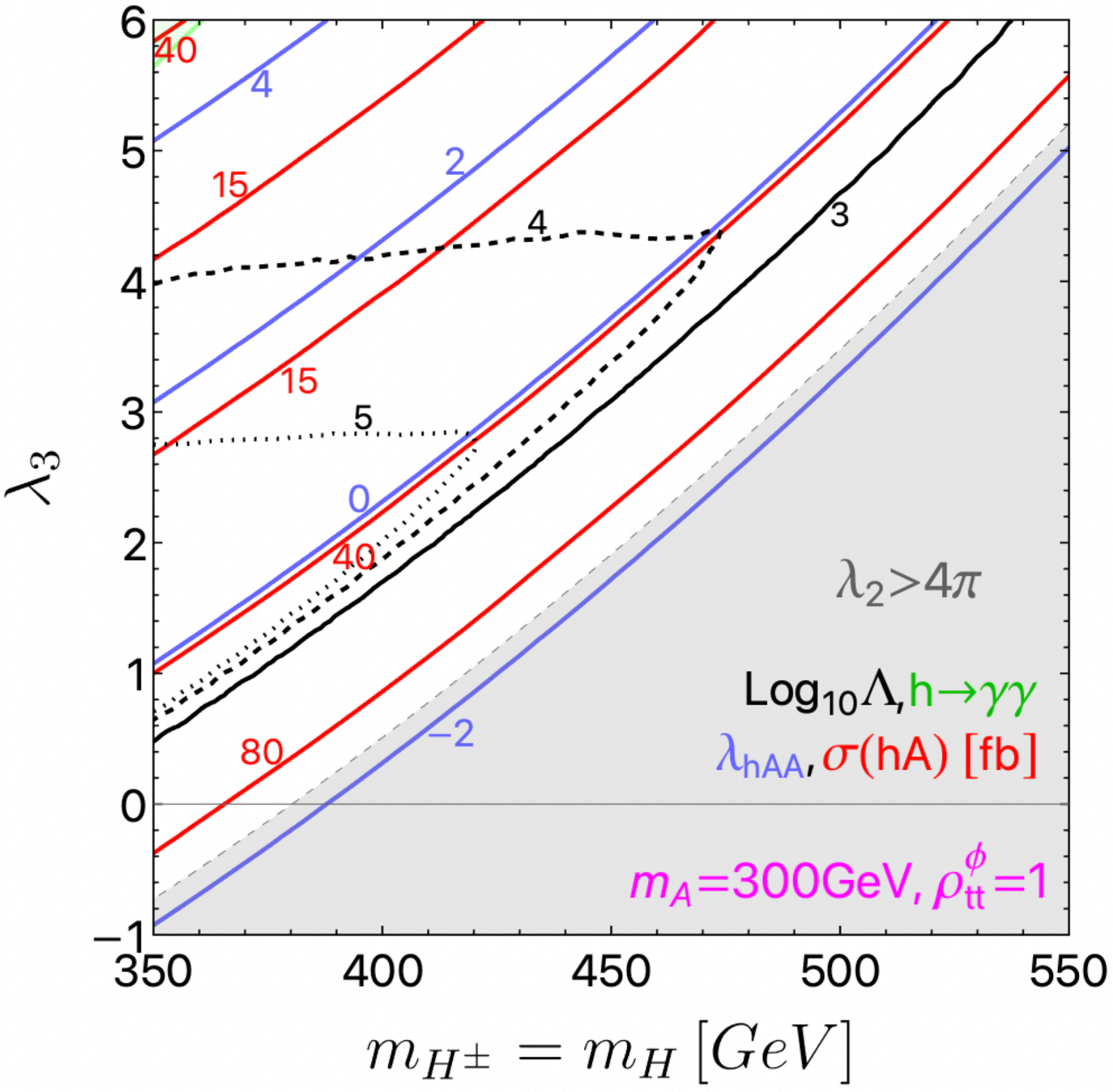}~~
\includegraphics[width=16.5em]{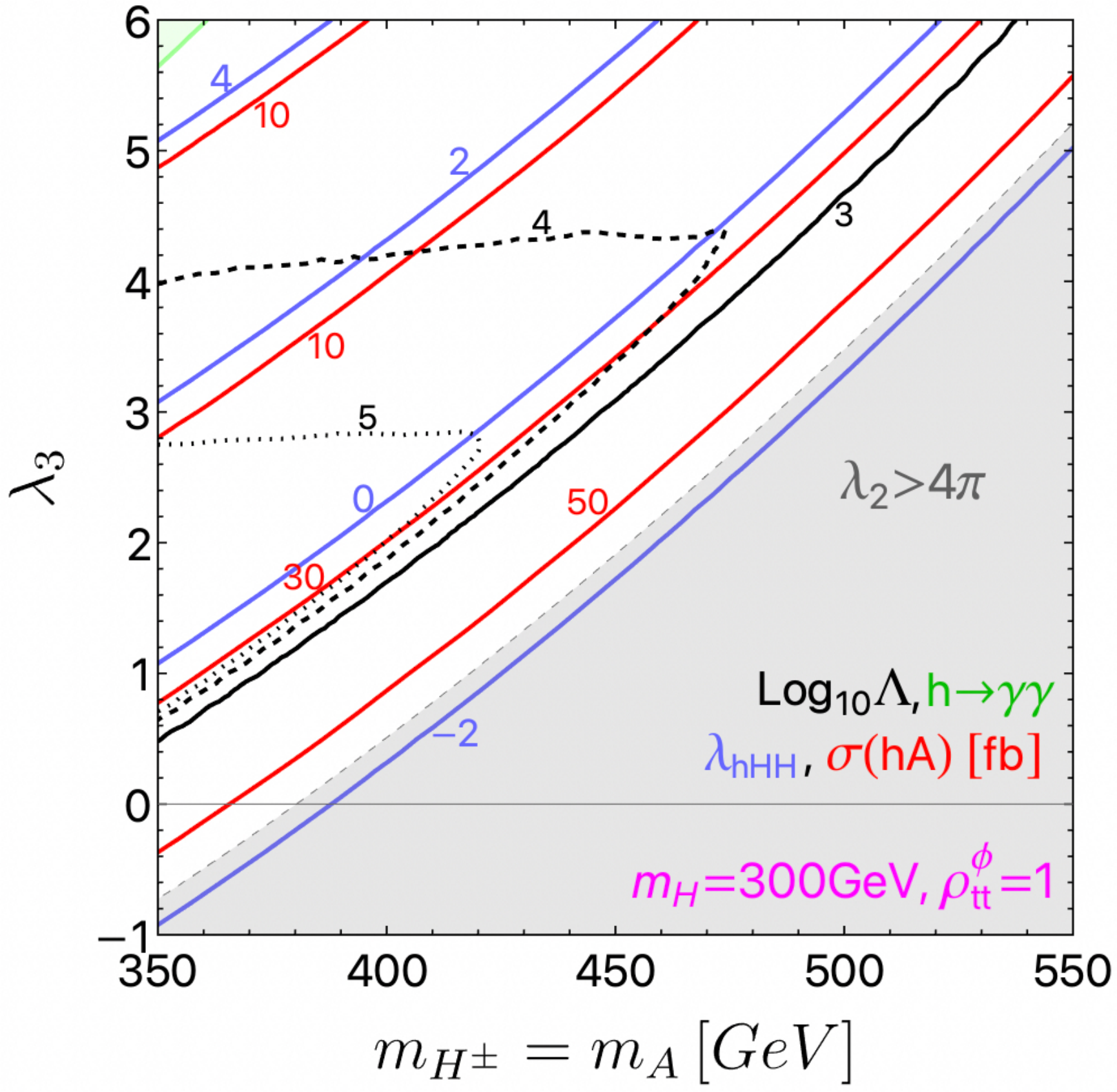}\\
\!\!\!\!\!
\includegraphics[width=16.5em]{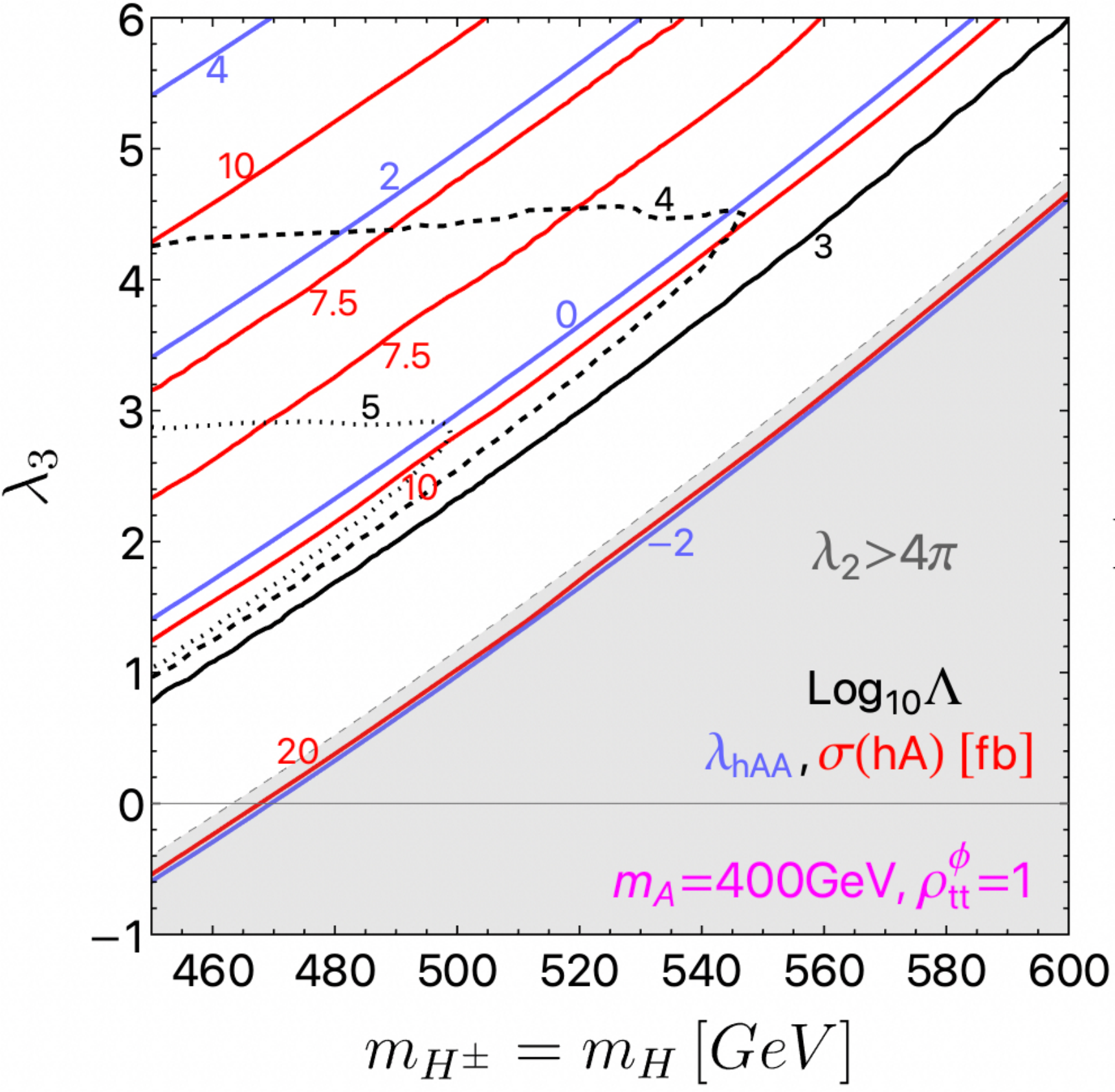}~~
\includegraphics[width=16.em]{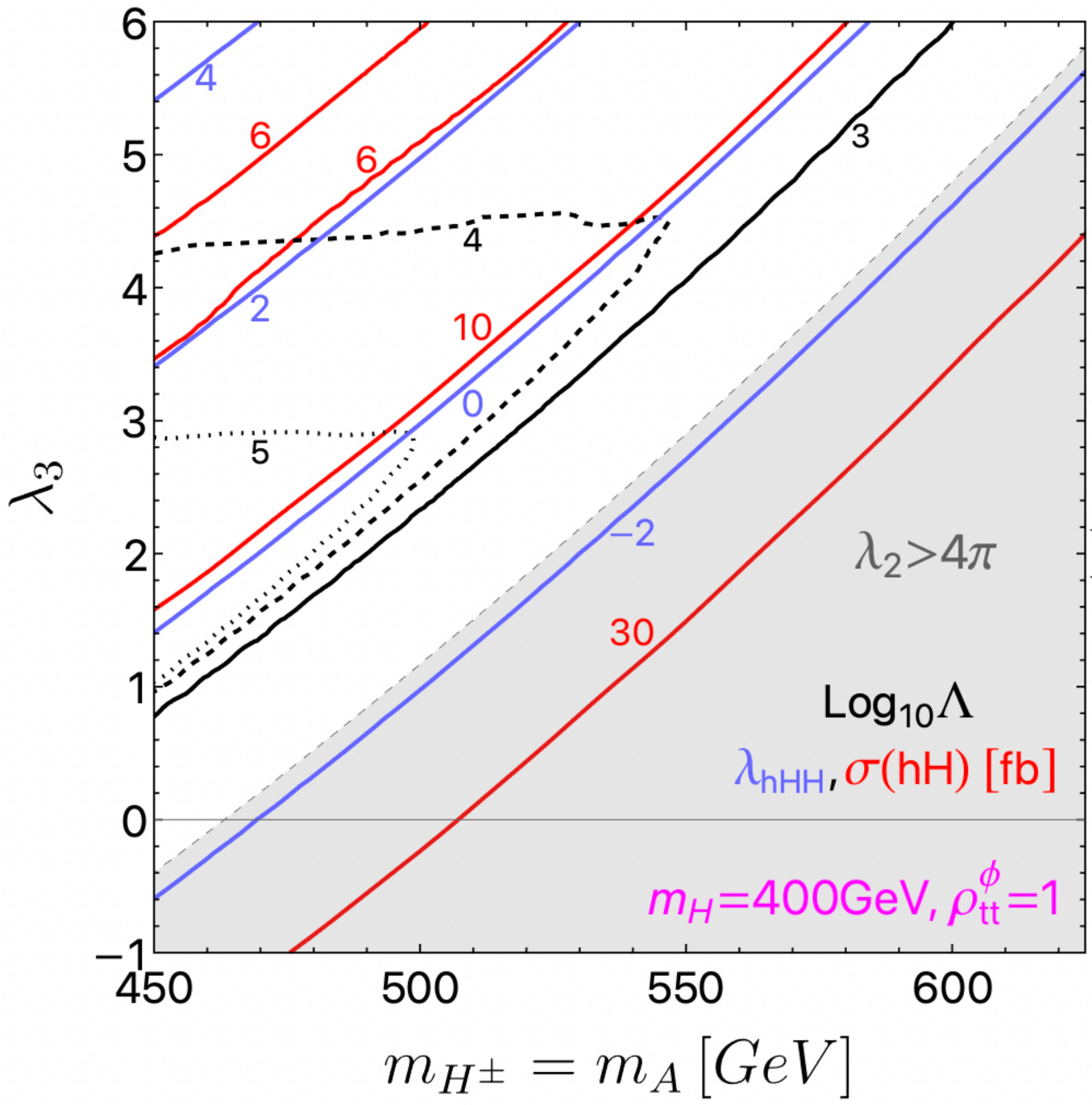}
\caption{\label{fig:Xs3}
The $h\phi_l$ production cross section is shown on the heavier scalar mass vs.\,$\lambda_3$ plane.
The left and right figures are for $hA$ and $hH$ productions, respectively.
The lighter scalar mass is fixed to be 250\,GeV\,(upper), 300\,GeV\,(middle), 400\,GeV\,(bottom).
See, the caption of Fig.~\ref{fig:Xs1} for the description of the constraints and other explanations.}
\end{center}
\end{figure}

%%%%%%%%%%%%%%%%%%%%%%%%%%%%%%%
In this Appendix,
we present figures that are not shown in the main text.
Figure~\ref{fig:Xs2} shows the production cross sections for $m_{\phi_l}=50$\,GeV (upper), 150\,GeV (middle) and 200\,GeV (bottom).
Figure~\ref{fig:Xs3} shows the results for $m_{\phi_l}=250$\,GeV (upper), 300\,GeV (middle) and 400\,GeV (bottom).
The color code is the same as Fig.~\ref{fig:Xs1} and for the description of the contours see the caption of Fig.~\ref{fig:Xs1} and the main text. 
For $m_{\phi_l}=50\,$GeV, the Higgs width bound gives a stringent constraint on $\lambda_{h\phi_l\phi_l}$ since $h \to \phi_l \phi_l$ is kinematically open.
Therefore, the allowed region from   the Higgs width bound must be almost degenerated to the line of $\lambda_{h\phi_l\phi_l}=0$. 
It is noted that the constraint from $h\to\gamma\gamma$ does not appear on the plane with $m_{\phi_l}=400\,$GeV.
As mentioned in Sec.\,\ref{sec:hphi_production}, these cross-section data are available in \cite{data}.
%%%%%%%%%%%%%%%%%%%%%%%%
%%%%%%%%%%%%%%%%%%%%%%%%

%%%%%%%%%%%%%%%%%%%%%%%%

%\FloatBarrier
%%%%%%%%%%%%%%%%%%%%%%%%%

%%%%%%%%%%%%%%%%%%%%%%%%%%%%%%%
\section{Parameter relations in the 2HDMs}
\label{sec:parameters}
%%%%%%%%%%%%%%%%%%%%%%%%%%%%%%%

For the sake of completeness, 
in this Appendix, we summarize the parameter relations between the general basis and the Higgs basis in the 2HDMs (see, \eg,  Refs.~\cite{Kanemura:2015ska,Iguro:2017ysu,Aiko:2020atr}).

The most general Higgs potential for the general 2HDM is given by 
\begin{align}
V = ~ & m_{11}^2 \Phi_1^{\dagger} \Phi_1+m_{22}^2 \Phi_2^{\dagger} \Phi_2-\left(m_{12}^2 \Phi_1^{\dagger} \Phi_2
+\text {h.c.}\right) \nonumber \\
& +\frac{1}{2} \lambda_1^G \left(\Phi_1^{\dagger} \Phi_1\right)^2+\frac{1}{2} \lambda_2^G \left(\Phi_2^{\dagger} \Phi_2\right)^2+\lambda_3^G \left(\Phi_1^{\dagger} \Phi_1\right)\left(\Phi_2^{\dagger} \Phi_2\right) +\lambda_4^G \left(\Phi_1^{\dagger} \Phi_2\right)\left(\Phi_2^{\dagger} \Phi_1\right) \nonumber  \\ 
& 
+\left\{ \frac{1}{2}\lambda_5^G \left(\Phi_1^{\dagger} \Phi_2\right)^2  + \left[\lambda_6^G \left(\Phi_1^{\dagger} \Phi_1\right)+\lambda_7^G \left(\Phi_2^{\dagger} \Phi_2\right)\right]\left(\Phi_1^{\dagger} \Phi_2\right)+\text {h.c.}\right\}\,,
\label{eq:potential_general}
\end{align}
with
\beq
\Phi_i &= \begin{pmatrix}
\omega_i^+ \\
\frac{1}{\sqrt{2}}\left( v_i + h_i + i z_i \right)
\end{pmatrix} \,, \quad (i=1,\,2)\,,
\eeq
where $v_i/\sqrt{2}$ is the VEV of $\Phi_i^0$, satisfying 
$ v = \sqrt{v_1^2 + v_2^2} = (\sqrt{2} G_F)^{- 1/2} \simeq 246\,\text{GeV}$ and
$
\tan \beta = {v_2}/{v_1}
$. Both $v_i$ can be taken to be real and positive values without losing generality. On the other hand, $m_{12}^2$ and $ \lambda_{5,6,7}^G$ are complex values in general.
When one imposes the softly-broken $\mathbbm{Z}_2$ symmetry
($\Phi_1 \to \Phi_1 , ~ \Phi_2 \to - \Phi_2$) to 
prohibit the flavor-changing neutral currents, $\lambda^G_{6,7} $ must be set to zero for any renormalization scale.

In the following, we assume the CP conservation for simplicity, and then $m_{12}^2$ and $ \lambda_{5,6,7}^G$ are real.
The stationary conditions for the general 2HDM potential are 
\beq
\begin{aligned}
m_{11}^2 - s^2_\beta M^2  + \frac{v^2}{2}
\left(c_\beta^2 \lambda_1^G  + s_\beta^2 \lambda^G_{345}   + 
3 s_\beta c_\beta \lambda_6^G  +  s^2_\beta \tan \beta \lambda_7^G \right) &= 0\,,\\
 m_{22}^2 - c^2_\beta M^2  + \frac{v^2}{2}
\left(s_\beta^2 \lambda_2^G  + c_\beta^2 \lambda^G_{345}   + 
  c_\beta^2 \cot \beta \lambda^G_6 + 3  s_\beta c_\beta  \lambda^G_7 \right)
& = 0\,, 
\label{eq:stationary}
\end{aligned}
\eeq
with
\beq
M^2 = \frac{m_{12}^2}{s_\beta c_\beta}\,,\quad \lambda^G_{345} = \lambda^G_3 + \lambda^G_4 + \lambda^G_5 \,,
\eeq
where  $s_\beta = \sin \beta$ and $c_\beta = \cos \beta$.

The Higgs basis in Eq.~\eqref{Eq:basis} is obtained by
\beq
\left(\begin{array}{l}
H_1 \\
H_2
\end{array}\right)=\left(\begin{array}{cc}
c_\beta & s_\beta\\
- s_\beta & c_\beta
\end{array}\right)\left(\begin{array}{l}
\Phi_1 \\
\Phi_2
\end{array}\right)\,.
\eeq
By matching the scalar potential in Eq.~\eqref{eq:potential_general} to the one in the Higgs basis~\eqref{Eq:potential}, we obtain the following parameter relations:
\begin{align}
M_{11}^{2} &= c^{2}_{\beta} m_{11}^{2}  + s^{2}_{\beta} m_{22}^{2}  - s_{2\beta} m_{12}^{2} \,, \\
M_{22}^{2} &= s^{2}_{\beta}  m_{11}^{2} + c^{2}_{\beta} m_{22}^{2}  +  s_{2\beta} m_{12}^{2} \,, \label{eq:M222}\\
M_{12}^{2} &= \frac{1}{2} s_{2\beta}\left(m_{11}^{2} - m^{2}_{22}\right) + c_{2\beta} m_{12}^{2} \,, \\
\lambda_{1} 
&=
c^{4}_{\beta} \lambda_{1}^G  + s^{4}_{\beta}\lambda_{2}^G  + \frac{1}{2} s^{2}_{2\beta} \lambda^G_{345} 
+ 2 s_{2 \beta} \left( c_{\beta}^2  \lambda_6^G 
+  s_{\beta}^2 \lambda_7^G \right)\,, \\
\lambda_{2} 
 &=s^{4}_{\beta}  \lambda_{1}^G + c^{4}_{\beta} \lambda_{2}^G  + \frac{1}{2} s^{2}_{2\beta}\lambda_{345}^G
-2 s_{ 2 \beta}  \left( s_{\beta}^2 \lambda_6^G +  c_{\beta}^2\lambda_7^G \right)  
\,, \\
\lambda_{3} 
 &= \frac{1}{4} s^{2}_{2\beta}\left(\lambda_{1}^G+ \lambda_{2}^G -2\lambda_{345}^G\right) + \lambda_{3}^G
 - \frac{1}{2} s_{4 \beta} \left(\lambda_6^G - \lambda_7^G \right) 
\,, \\
\lambda_{4} 
&= 
\frac{1}{4} s^{2}_{2\beta}\left(\lambda_{1}^G+ \lambda_{2}^G -2\lambda_{345}^G\right) + \lambda_{4}^G
 - \frac{1}{2} s_{4 \beta} \left(\lambda_6^G - \lambda_7^G \right) 
\,, \\
\lambda_{5} 
&= 
\frac{1}{4} s^{2}_{2\beta}\left(\lambda_{1}^G+ \lambda_{2}^G -2\lambda_{345}^G\right) + \lambda_{5}^G
 - \frac{1}{2} s_{4 \beta} \left(\lambda_6^G - \lambda_7^G \right) 
 \,, \\
\lambda_{6} 
&= -\frac{1}{2} s_{2\beta}\left(c^{2}_{\beta}\lambda^G_{1} -  s^{2}_{\beta} \lambda^G_{2} -  c_{2\beta}\lambda^G_{345} \right)
+ c_\beta^2 \left( 1 - 4 s_\beta^2 \right) \lambda_6^G + s_\beta^2 \left( -1 + 4 c_{\beta}^2\right) \lambda_7^G
\,, \\
\lambda_{7} 
&= -\frac{1}{2} s_{2\beta}\left( s^{2}_{\beta} \lambda_{1}^G- c^{2}_{\beta}\lambda_{2}^G + c_{2\beta}\lambda_{345}^G\right) 
 + s_\beta^2\left( -1 + 4 c_{\beta}^2\right) \lambda_6^G + c_\beta^2 \left( 1 - 4 s_\beta^2 \right) \lambda_7^G
\,.
\end{align}
Using the stationary equations \eqref{eq:stationary}, $M_{11}^2$ and $M_{12}^2$ can be represented as 
\beq
M_{11}^2 &=
-\frac{v^2}{2}\left( 
 c_\beta^4 \lambda_1^G+ s_\beta^4\lambda_2^G  + \frac{1}{2} s_{2\beta}^2 \lambda_{345}^G + 4 c_\beta^3 s_\beta \lambda_6^G + 4 s_\beta^3 c_\beta \lambda_7^G \right)\nonumber \\
& = - \frac{1}{2} \lambda_1 v^2\,,\\
M_{12}^2 &= 
\frac{v^2}{2}\left[ - \frac{1}{2} s_{2\beta}\left( c_\beta^2  \lambda_1^G- s_\beta^2 \lambda_2^G - c_{2\beta} \lambda_{345}^G\right) + c_\beta^2 \left( 1 - 4 s_\beta^2 \right)\lambda_6^G + s_{\beta}^2\left( -1 + 4 c_{\beta}^2\right)  \lambda_7^G\right] \nonumber \\
& = \frac{1}{2} \lambda_6 v^2\,.
\eeq

When one imposes the approximate Higgs alignment condition ($\lambda_6 \simeq 0$) in order to 
avoid the experimental bounds from measurements of the Higgs signal strengths, the scalar potential is given as
\begin{align}
\begin{aligned}
  V\simeq & - \frac{1}{2} \lambda_1 v^2 H_1^\dagger H_1+M_{22}^2 H_2^\dagger H_2 \\
&+\frac{\lambda_1}{2}(H_1^\dagger H_1)^2+\frac{\lambda_2}{2}(H_2^\dagger H_2)^2+\lambda_3(H_1^\dagger H_1)(H_2^\dagger H_2)
+\lambda_4 (H_1^\dagger H_2)(H_2^\dagger H_1) \\
&+
\left[\frac{\lambda_5}{2}(H_1^\dagger H_2)^2+ \lambda_7 (H_2^\dagger H_2) (H_1^\dagger H_2)+{\rm h.c.}\right]\,.
   \end{aligned}
\end{align}
The scalar boson masses are determined as 
\beq
\begin{aligned}
m_h^2 &\simeq \lambda_1 v^2\,, \quad & m_H^2&\simeq M_{22}^2 + \frac{1}{2} \lambda_{345} v^2\,,\\
m_A^2 &= M_{22}^2 + \frac{1}{2} \left( \lambda_3 + \lambda_4 - \lambda_5\right) v^2\,, \quad & m_{H^\pm}^2 & = M_{22}^2 + \frac{1}{2}\lambda_3 v^2\,.
\end{aligned}
\eeq
Given four scalar boson masses, the remaining degrees of freedom in this model are three (\eg, $\lambda_{2,3,7}$), and the value of $\lambda_7$ is irrelevant to our study except for an indirect effect on the perturbative unitarity bound. 
As we discussed below Eq.~\eqref{eq:BFB}, 
$\lambda_2$ gives only a relevant effect on the vacuum stability condition. 
As a consequence, the production cross-section of $gg\to h \phi_l$ can be determined only by  $\lambda_3$ and the scalar masses.

When one imposes the (softly-broken) $\mathbbm{Z}_2$ symmetry,
$\lambda_{6,7}^G$ are forbidden. This condition provides the following parameter relations in the Higgs basis:
\beq
\lambda_6^G 
&= \frac{1}{2} s_{2 \beta} \left( c_\beta^2 \lambda_1 - s_\beta^2 \lambda_2 - c_{2 \beta} \lambda_{345}\right) + c_\beta c_{3 \beta} \lambda_6 + s_\beta s_{3 \beta} \lambda_7 =0\,, \\
\lambda_7^G 
&= \frac{1}{2} s_{2 \beta} \left( s_\beta^2 \lambda_1 - c_\beta^2 \lambda_2 + c_{2 \beta} \lambda_{345}\right) + s_\beta s_{3 \beta} \lambda_6 + c_\beta c_{3 \beta} \lambda_7 =0\,,
\eeq
which further restrict the parameters.
Note that in this paper we do not impose these conditions.

%%%%%%%%%%%%%%%%%%%%%%%%%%%%%%%%%%%%%%%%%%%%%%%%%%%%%%%%
%%%%%%%%%%%%%%%%%%%%%%%%%%%%%%%%%%%%%%%%%%%%%%%%%%%%%%%%

%\bibliographystyle{apsrev4-1_title}
%\bibliographystyle{JHEP}
\bibliographystyle{utphys28mod}
{\footnotesize
\bibliography{ref}
}
\end{document}